\documentclass[10pt,twocolumn,aps,tightenlines,floats,amsmath,amssymb,prd,floatfix,longbibliography,superscriptaddress,notitlepage,nofootinbib]{revtex4-2}

\usepackage[utf8]{inputenc}
\usepackage[T1]{fontenc}
\usepackage{wasysym}
\usepackage{amsmath,amssymb,amsfonts,amsthm,mathrsfs}
\usepackage{graphicx,wrapfig}
\usepackage{enumerate}
\usepackage{enumitem}
\usepackage{placeins}
\usepackage{xcolor}
\usepackage{booktabs}
\definecolor{lightblue}{RGB}{51,102,204} 
\definecolor{lightred}{RGB}{191,60,45} 
\usepackage{physics}
\usepackage{slashed}
\usepackage{mathtools}

\usepackage{orcidlink}
\usepackage{bbm}
\usepackage[caption=false]{subfig}

\usepackage{hyperref}
\hypersetup{
	bookmarks=true,         
	unicode=false,          
	pdftoolbar=true,        
	pdfmenubar=true,        
	pdffitwindow=false,     
	pdfstartview={FitH},    
	pdftitle={Causal spinfoam vertex for Lorentzian quantum gravity},    
	pdfauthor={Mauricio Gamonal, Eugenio Bianchi, Chaosong Chen},     
	pdfkeywords={causality, loop quantum gravity, quantum gravity}, 
	pdfnewwindow=true,      
	colorlinks=true,       
	linkcolor=lightred,          
	citecolor=lightblue,        
	filecolor=cyan,         
	urlcolor=lightblue        
}

\newcommand{\ii}{\mathrm{i}}
\newcommand{\ee}{\mathrm{e}}
\newcommand{\CC}{\mathbb{C}}
\newcommand{\id}{\mathbbm{1}}

\newcommand{\outcomment}[1]{}

\begin{document}

\title{\Large Causal spinfoam vertex for 4d Lorentzian quantum gravity}

\author{Eugenio Bianchi\,\orcidlink{0000-0001-7847-9929}}
\email{ebianchi@psu.edu}
\author{Chaosong Chen\,\orcidlink{0009-0003-6722-2126}}
\email{cchen@psu.edu}
\author{Mauricio Gamonal\,\orcidlink{0000-0002-0677-4926}}
\email{mgamonal@psu.edu}

\affiliation{Institute for Gravitation and the Cosmos, The Pennsylvania State University, University Park, Pennsylvania 16802, USA}\affiliation{Department of Physics, The Pennsylvania State University, University Park, Pennsylvania 16802, USA}

\date{\today}

\begin{abstract}
We introduce a new causal spinfoam vertex for $4$d Lorentzian quantum gravity. The causal data are encoded in Toller $T$-matrices, which add to Wigner $D$-matrices $T^{(+)}+T^{(-)}=D$, and for which we provide a Feynman $\mathrm{i}\varepsilon$ representation. We discuss how the Toller poles cancel in the EPRL vertex, how the Livine-Oriti model is obtained in the Barrett-Crane limit, and how spinfoam causal data are distinct from Regge causal data. In the large-spin limit, we show that only Lorentzian Regge geometries with causal data compatible with the spinfoam data are selected, resulting in a single exponential $\exp(+\mathrm{i}\, S_{\mathrm{Regge}}/\hbar)$ and a new form of causal rigidity.

\end{abstract}

\maketitle




\section{Introduction}
\label{sec:Introduction}

Spinfoams provide the covariant path-integral formulation of loop quantum gravity (LQG) \cite{Rovelli:2014ssa,Ashtekar:2021kfp}. The elementary transitions are encoded in the spinfoam vertex amplitude \cite{Perez:2012wv,Engle:2023qsu,Livine:2024hhc}. The Engle–Pereira–Rovelli–Livine (EPRL) model~\cite{Engle:2007wy} defines a spinfoam vertex amplitude for $4$d Lorentzian quantum gravity. In this paper we introduce a new spinfoam vertex which provides a causal version of the EPRL model with a fixed causal structure.

Causal structures in the spinfoam path integral \cite{Reisenberger:1996pu,Markopoulou:1997wi,Bianchi:2021ric} have been studied extensively \cite{Markopoulou:1997hu,Markopoulou:1999cz,Gupta:1999cp,Livine:2002rh,Oriti:2004mu,Pfeiffer:2002ic,Hawkins:2003vc,Freidel:2005bb,Oriti:2005jr,Oriti:2006wq,Livine:2006xc,Rovelli:2012yy,Bianchi:2012nk,Oriti:2013aqa,Immirzi:2013rka,Cortes:2014oka,Wieland:2014nka,Immirzi:2016nnz,Finocchiaro:2018hks,Jercher:2022mky,Simao:2024don,Oriti:2025uad}, but a causal version of the EPRL model has remained elusive so far. Motivated by the causal-set approach \cite{Bombelli:1987aa,Sorkin:2003bx,Dowker:2005tz,Surya:2019ndm} and by the construction of the causal propagator \cite{Teitelboim:1981ua}, Livine and Oriti identified the causal version \cite{Livine:2002rh} of the Barrett-Crane model \cite{Barrett:1999qw}. Here we follow a similar strategy. The new ingredient is the use of Toller $T$-matrices, defined in the literature on harmonic analysis on the Lorentz group \cite{Ruhl:1970lor} by their pole structure, the Toller poles \cite{Toller:1968gr,Toller:1968pole,Sciarrino:1967,Taylor:1967xsb,Jones:1969cd}. Schematically, they satisfy the relation
\begin{equation}
    T^{(+)}+T^{(-)}=D\,,
\end{equation}
where $D$ is the Wigner $D$-matrix for the Lorentz group which appears in the EPRL model. 
We show that the Toller matrices can be equivalently formulated in terms of a Feynman $\ii\varepsilon$ prescription, which naturally encodes the causal structure in spinfoams (Sec.~\ref{sec:Wigner-Toller}). We discuss how the Toller poles cancel in the EPRL model, and show that in the Barrett-Crane limit we recover the Livine-Oriti causal model (Sec.~\ref{sec:Relation-EPRL}). In a companion paper \cite{Bianchi:2025Toller}, we provide further details on the Toller $T$-matrices.

In the large-spin limit, for a boundary state peaked on the geometry of a non-degenerate Lorentzian $4$-simplex, the EPRL amplitude reproduces the cosine of the Regge action \cite{Regge:1961px}. This result is based on a saddle-point analysis \cite{Barrett:2009gg,Barrett:2009mw,Han:2013hna,Dona:2020xzv,Dona:2020yao,Dona:2022hgr,Han:2020fil} and confirmed by numerical investigations \cite{Speziale:2016axj,Dona:2018nev,Dona:2019dkf,Dona:2020tvv,Gozzini:2021kbt,Dona:2022dxs,Dona:2022yyn}. In \cite{Bianchi:2021ric}, it was argued that a fixed causal structure can be encoded in the EPRL model as the insertion of a step function in the coherent state representation. The effect of this \emph{ad hoc} modification is to restrict the region of integration, resulting in a single saddle point which reproduces the behavior of Engle's proper vertex \cite{Engle:2011un,Engle:2012yg,Engle:2015mra,Engle:2015zqa}, as argued by Immirzi in \cite{Immirzi:2016nnz}. Here we show that the construction in terms of Toller $T$-matrices encodes naturally the causal structure and reproduces these properties in the large-spin limit (Sec.~\ref{sec3:Asymptotics}).

\section{Causal vertex and Toller matrices}
\label{sec:Wigner-Toller}

\begin{figure*}[t]
  \centering
  \subfloat[Combinatorial causal data: $\sigma_a$]{
\includegraphics[width=0.35\textwidth]{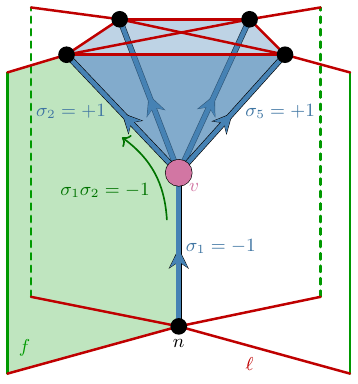}
    \label{fig:Causal-Spinfoam}
  }\hspace{3em}
  \subfloat[Regge causal data: $s_a$]{
\includegraphics[width=0.5\textwidth]{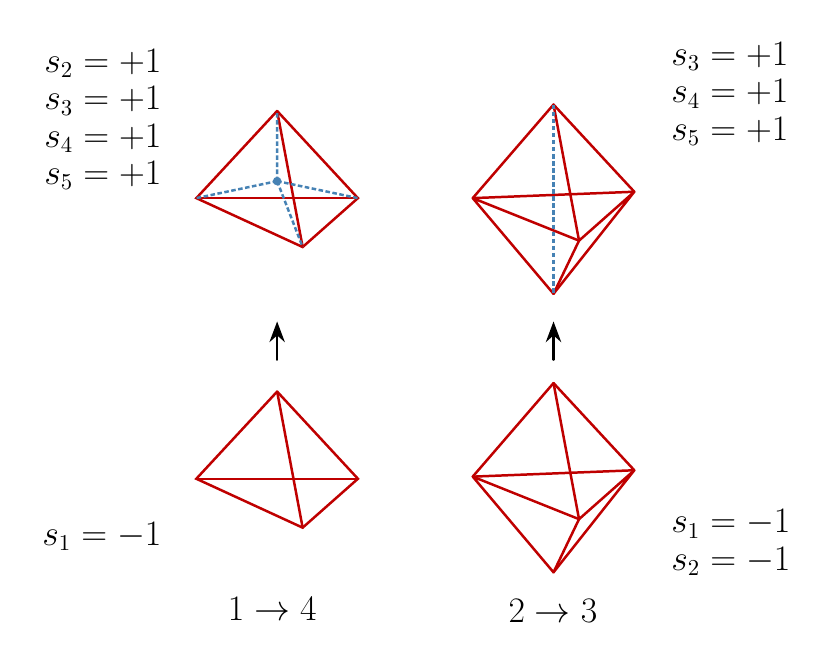}
    \label{fig:Causal-Tetrahedra}
  }
  \caption{\emph{Panel} (a): The combinatorial causal data $\sigma_a$ are associated with each oriented edge (blue arrows) ingoing or outgoing from a vertex $v$ (purple dot). The figure depicts a $1\to 4$ transition from one spin-network node $n$ to four nodes (black dots). The wedge $(12)$ bounded by two edges (green plane) carries causal data $\sigma_1 \sigma_2=-1$. \emph{Panel} (b): The Regge causal data $s_a$ are associated with the $4$-normals to the spacelike tetrahedra bounding a Lorentzian $4$-simplex. The figure depicts the Regge transitions $1\to 4$ and $2\to 3$. Of the two, only the $1\to 4$ transition---with one tetrahedron on the past boundary and four on the future boundary of the $4$-simplex---is compatible with the combinatorial data depicted in panel (a), as it satisfies the causal rigidity condition $\sigma_a \sigma_b=s_a s_b$, for all couples $(ab)$.}
  \label{fig:Spinfoam}
\end{figure*}

The spinfoam path integral is defined by a sum over $2$-complexes (combinatorial objects consisting of vertices $v$, edges $e$, and faces $f$), together with a sum over spins $j_f$ associated with faces and intertwiners $i_e$ associated with edges (see \cite{Perez:2012wv,Engle:2023qsu,Livine:2024hhc} for recent reviews). A \emph{causal} spinfoam path integral also includes an additional structure: a sum over the orientation of edges 
\cite{Markopoulou:1997hu,Markopoulou:1997wi,Markopoulou:1999cz,Gupta:1999cp,Livine:2002rh, Bianchi:2021ric,Oriti:2004mu}. This orientation is required to determine a partial order of the vertices of the $2$-complex, which then form a causal set \cite{Bombelli:1987aa,Sorkin:2003bx,Dowker:2005tz,Surya:2019ndm}. Here, we consider a single vertex $v$ dual to a $4$-simplex and denote its edges by $a=1,\ldots,5$. The causal orientation of each edge is given by a variable $\sigma_a=\pm 1$,
\begin{equation}
\sigma_a= 
\begin{cases}
 -1  & \textrm{for an ingoing edge},   \\[.5em]
+1  &  \textrm{for an outgoing edge}.
\end{cases}
\end{equation}
A wedge $(ab)$ is defined by two edges $a$ and $b$ at the vertex $v$. Two edges form a co-chronal wedge if $\sigma_a\sigma_b=+1$ (also called a thick wedge), or an anti-chronal wedge if $\sigma_a\sigma_b=-1$ (also called a thin wedge), (see Fig.~\ref{fig:Causal-Spinfoam}). 

Note that, reversing the orientation of all edges at a vertex, $\sigma_a\to -\sigma_a$ for all $a$, does not change the causal relation of the wedges, $\sigma_a \sigma_b\to \sigma_a\sigma_b$. Therefore the inequivalent transitions fall into three classes: $0\leftrightarrow 5$, $1\leftrightarrow 4$, and $2\leftrightarrow 3$ transitions.

Note also that this notion of causal structure is purely combinatorial and should not be confused with the geometric notion of causal structure in a Regge geometry, where the future-pointing or past-pointing $4$-normals to spacelike tetrahedra on the boundary of a Lorentzian $4$-simplex are denoted by $s_a=\pm1$, (see Fig.~\ref{fig:Causal-Tetrahedra}). A relation between the two arises only at the semiclassical level, as discussed later.

In the spinfoam path integral, the sum over the combinatorial structures is weighted by a vertex amplitude, which can be understood as a linear functional $\langle A_v|$ acting on boundary states. The EPRL model \cite{Engle:2007wy} is defined by a vertex amplitude $\langle A_v^{\mathrm{EPRL}}|$ built from the Wigner $D$-matrices $D^{(\rho,k)}_{jm\, ln}(g)$ associated with each wedge $(ab)$. These matrices provide a unitary representation of the Lorentz group elements $g\in SL(2,\CC)$, given by the principal series $(\rho,k)\in (\mathbb{R},\mathbb{Z}/2)$. In the EPRL model, only the  $\gamma$-simple representations $(\rho,k)=(\gamma\, j_{ab},j_{ab})$ appear, where $\gamma$ is the Barbero-Immirzi parameter, $j_{ab}$ is the spin associated with a wedge $(ab)$, and the Lorentz group element $g_{ab}= g_b^{-1}g_a^{\vphantom{-1}}$ represents the parallel transport along the edges $a$ and $b$ that form the wedge. Note that this model is independent of the combinatorial causal data $\sigma_a$. In this paper we introduce a causal vertex amplitude for $4$d Lorentzian quantum gravity, which depends explicitly on the causal data $\sigma_a$. Its definition requires a new ingredient: the Toller $T$-matrices \cite{Toller:1968gr,Toller:1968pole,Sciarrino:1967}. These are polynomially-bounded functions on $SL(2,\mathbb{C})$ which can be defined via a Feynman $\ii  \varepsilon$ prescription as \cite{Bianchi:2025Toller}:
\begin{widetext}
\begin{equation}
T^{(\pm,\,\rho,\,k)}_{jm\,ln}(g)=\lim_{\varepsilon \to 0^+}\int_{-\infty}^{+\infty}\frac{\dd{\tilde{\rho}}}{2\pi\ii}\;\frac{\pm1}{\tilde{\rho}-\rho \mp\ii\, \varepsilon} \; \frac{\Gamma (-j-\ii \rho) \Gamma (l-\ii \tilde{\rho} +1)}{\Gamma (-j-\ii \tilde{\rho} ) \Gamma (l-\ii \rho+1)}\, D^{(\tilde{\rho},k)}_{jm\, ln}(g)\,,
\label{eq:Feynman-ieps}
\end{equation}
\end{widetext}
where $\Gamma(x)$ is the Gamma function. To each wedge $(ab)$ in a spinfoam vertex, we associate a Toller $T$-matrix with sign $\pm$ determined by the wedge orientation $\kappa_{ab}=\sigma_a \sigma_b$ induced by the orientations $\sigma_a$ and $\sigma_b$ of the two edges forming the wedge.  

\begin{figure*}[t]
	\centering
\includegraphics[width=\textwidth]{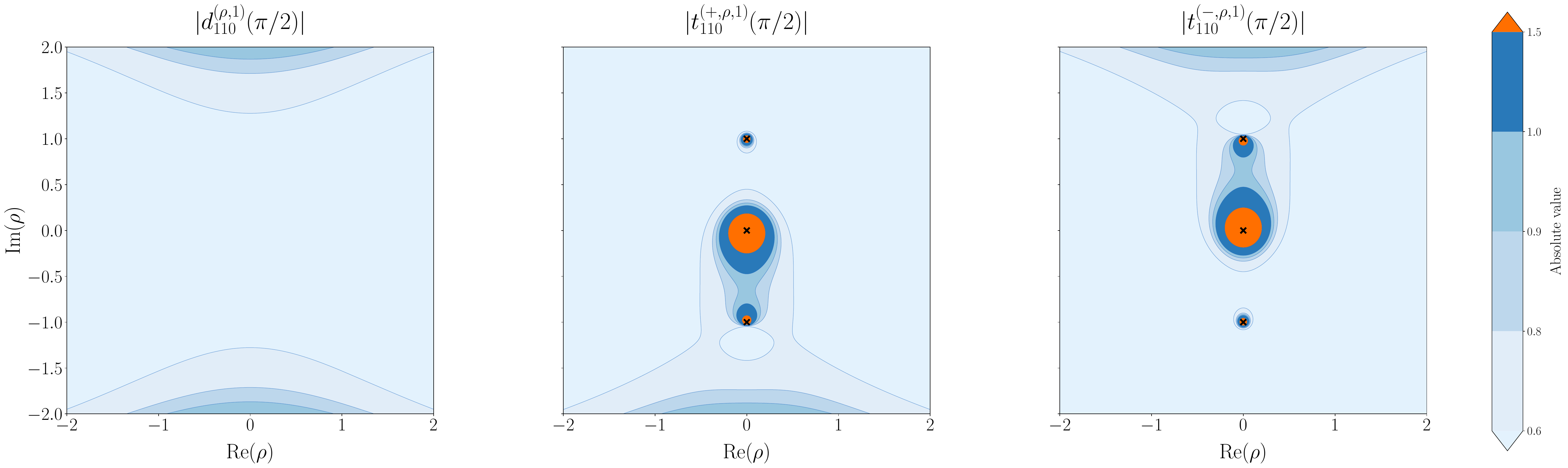}
	\caption{Contours in the complex $\rho$-plane of the absolute value of the reduced Wigner and Toller matrices. For illustration, we display the case $k=j=l=1$, $m=0$, and $\beta=\pi/2$; their explicit analytic forms are provided in  Appendix~\ref{app:BoosterFunctions}. The markers $\times$ indicate the positions of the Toller poles at $\ii \rho = -1, 0,+1$. The figure highlights the analytic structure, asymptotic fall-off, and matching conditions that characterize these functions.}
	\label{fig:Contour-Plot}
\end{figure*}

For a spin-network boundary state $ \ket{\Psi_{j_{ab},i_a}} = \sum_{m_{ab}} (i_a)_{m_{ab}}  \ket{\Psi_{j_{ab},m_{ab}}}$, defined by $10$ spins $j_{ab}$ and $5$ interwiners $(i_a)_{m_{ab}}$ (with $m_{ab}$ magnetic numbers), replacing Wigner $D$-matrices with Toller $T$-matrices in the definition of the EPRL model, one obtains the vertex amplitude
\begin{equation}
\label{eq:causal-vertex-def}
\langle A^{(\sigma_a \sigma_b)}_v |\Psi_{j_{ab},m_{ab}}\rangle \! =\! \int\!\! \prod_{a=2}^{5} \dd{g_a} 
\prod_{ab} T^{(\sigma_a\sigma_b,\,\gamma j_{ab},j_{ab})}_{j_{ab} \,m_{ba}\,j_{ab} \, m_{ab}} (g_b^{-1}g_a^{\vphantom{-1}})\,,
\end{equation}
where the product over
$a b$ runs from $1\leq a< b \leq 5$, and we have gauge-fixed the $SL(2,\mathbb{C})$ invariance of the vertex to $g_1=\id$ by introducing a $\delta(g_1)$, \cite{Engle:2008ev, Dona:2022hgr}. Eq.~\eqref{eq:causal-vertex-def} defines the vertex amplitude $\langle A^{(\sigma_a \sigma_b)}_v|$ with fixed causal structure $\sigma_a$, which we propose and study in this paper.

\section{Relation to the EPRL model}
\label{sec:Relation-EPRL}

The Feynman $\ii \varepsilon$ formula introduced in Eq.~\eqref{eq:Feynman-ieps} encodes completely the properties of the Toller $T$-matrices, and is equivalent to their definition in terms of Toller poles discussed in R\"uhl's textbook \cite{Ruhl:1970lor} and in Appendix \ref{app:FullExpressions}. In particular, there are two properties that are immediate to prove from \eqref{eq:Feynman-ieps}, which allow us to relate the causal vertex amplitude \eqref{eq:causal-vertex-def} to the EPRL vertex amplitude. 

The first property is the additive relation:
\begin{equation}
\label{eq:Additive-Property}
T^{(+,\rho,k)}_{jm\,l n}(g) + T^{(-,\rho,k)}_{jm\,l n}(g) = D^{(\rho,k)}_{jm\,l n}(g)\, .
\end{equation}
This property follows immediately from the principal-value relation $\frac{1}{x\pm\ii \varepsilon}=\mathcal{P}(\frac{1}{x})\,\mp\ii \pi \delta(x)$. While the Wigner $D$-matrices are unitary irreducible representations of $SL(2,\CC)$, the Toller $T$-matrices are functions over $SL(2,\CC)$ which are polynomially bounded in $\tr(g g^\dagger)$ and are fully determined by the expansion \eqref{eq:Feynman-ieps}. As a result of \eqref{eq:Additive-Property}, the EPRL vertex amplitude can be expressed as an \emph{unconstrained} sum over wedge signs $\kappa_{ab}=\pm 1$,
\begin{equation}
    \langle A_v^{\mathrm{EPRL}}|\,=\sum_{\kappa_{ab}=\pm 1}\langle A_v^{({\kappa_{ab}})}|\,,
\end{equation}
where $\langle A_v^{({\kappa_{ab}})}|$ is defined as in Eq.~\eqref{eq:causal-vertex-def} in terms of $T^{(\kappa_{ab},\,\gamma j_{ab},j_{ab})}_{j_{ab} \,m_{ba}\,j_{ab} \, m_{ab}} (g_b^{-1}g_a^{\vphantom{-1}})$. On the other hand, the sum over causal structures $\sigma_{a}=\pm 1$ results in a constrained sum over wedge signs. Therefore we have that a sum over causal structures does not reproduce the EPRL vertex amplitude, $\sum_{\sigma_a=\pm 1}\langle A_v^{({\sigma_a \sigma_b})}|\neq \langle A_v^{\mathrm{EPRL}}|$. The consequences of this distinction are discussed later.

The second property of Toller $T$-matrices that follows directly from the Feynman $\ii \varepsilon$ formula \eqref{eq:Feynman-ieps} is their expression in the coherent state basis \eqref{eq:Toller-coherent-basis}. This formula will allow us to compute the semiclassical limit of the causal vertex, using techniques analogous to the ones adopted for the EPRL model.

In detailed calculations and numerical implementations of the vertex amplitude \cite{Speziale:2016axj,Dona:2018nev,Dona:2019dkf,Dona:2020tvv,Gozzini:2021kbt,Dona:2022dxs,Dona:2022yyn}, it is useful to have an explicit expression for the Toller $T$-matrices for a pure boost $\ee^{\beta \frac{\sigma_z}{2}}$, with $\beta> 0$. Using the Cartan decomposition $g=U_1\, \ee^{\beta \frac{\sigma_z}{2}}\, U_2$, with $U_1,U_2\in SU(2)$, we have $D^{(\rho,k)}_{jm\,ln}(g) = \sum_{p} D^{(j)}_{mp} (U_1) \, d^{(\rho,k)}_{jlp}(\beta) \, D^{(l)}_{pn}(U_2) $, and similarly for the Toller $T$-matrices,
\begin{equation}
\label{eq:T-Cartan}
T^{(\pm,\,\rho,k)}_{jm\,ln}(g)=\!\!\!\!\!\!\sum_{p=-\min(j,l)}^{\min(j,l)} \!\!\!\!\! D^{(j)}_{mp}(U_1)\,t^{(\pm,\,\rho,k)}_{jlp}(\beta)\,D^{(l)}_{pn}(U_2)\, ,
\end{equation}
where $D^{(j)}_{mp}(U)$ is the $SU(2)$ Wigner matrix, and $\beta$ is the boost parameter. The reduced Toller $t$-matrices $t^{(\pm,\,\rho,k)}_{jlm}(\beta)$ defined in this way coincide with the ones discussed extensively in Sec.~4.5 of R\"uhl's textbook \cite{Ruhl:1970lor}, where they are denoted by $e$ and $f$. While the reduced Wigner $d$-matrices $d^{(\rho,k)}_{jlm}(\beta)$ are analytic in the complex $\rho$-plane, the reduced Toller $t$-matrices $t^{(\pm,\,\rho,k)}_{jlm}(\beta)$ are meromorphic with simple poles located at $\ii \, \rho = -j,\cdots , l$, which are also commonly known as Toller poles in the context of scattering theory \cite{Sciarrino:1967,Taylor:1967xsb,Toller:1968gr,Jones:1969cd,Toller:1968pole}, hence the terminology. These analyticity properties are fully captured by the Feynman $\ii \varepsilon$ formula \eqref{eq:Feynman-ieps} as summarized in Appendix \ref{app:FullExpressions}. For concreteness, we report explicit expressions for the $\gamma$-simple representation, where the reduced Wigner $d$-matrices $d^{(\gamma \, j,j)}_{jjm}(\beta)$ (see \cite{Speziale:2016axj}) and the reduced Toller $t$-matrices  $t^{(\pm,\,\gamma j,\, j)}_{jjm}(\beta)$ can be expressed in terms of hypergeometric functions 
${}_2F_1(
{\scriptsize
\begin{matrix}
	a, b \\[-.2em]
	c
\end{matrix}
}\,;z)$ as 
\begin{widetext}
\begin{align}
d^{(\gamma \, j,j)}_{jjm}(\beta) &\:=\,
\ee^{-(j - \ii \gamma \, j + m +1)\beta} \; {}_2F_1\!\left(
		 \begin{matrix}
		   j + m + 1,\; j+1 - \ii\, \gamma \, j \\
		    2j + 2
		 \end{matrix}
		 ;\, 1-\ee^{-2\beta}
	   \right)\, \label{eq:d-gamma-simple},\\[1em]
t^{(\pm,\,\gamma j,\, j)}_{jjm}(\beta) &\:=\, \ee^{-(j \,\mp\, \ii \,\gamma \, j \,\pm \,m \,+1)\beta} \; \frac{\Gamma(2j+2)\Gamma(\pm \,\ii\, \gamma\,j \mp m)}{\Gamma(j\,\mp\, m+1)\Gamma(j+1\pm \,\ii \,\gamma \, j)} \; {}_2F_1\!\left(
		 \begin{matrix}
		   j \pm m + 1,\; j+1 \mp \,\ii \,\gamma \, j \\
		    1\pm m \,\mp \,\ii \,\gamma \, j
		 \end{matrix}
		 ;\, \ee^{-2\beta}
	   \right)\,. \label{eq:t-gamma-simple}
\end{align}
\end{widetext}
The general expressions for arbitrary values of $\rho$, $k$, $j$ and $l$ are provided in \eqref{eq:APP-t-plus} and \eqref{eq:APP-t-minus}. Other relevant expressions for the Toller $t$-matrices in terms of an analytic continuation of boosts, and their relation to the Wick rotation from $Spin(4)$ to $SL(2,\CC)$ \cite{Dona:2021ldn}   are discussed in detail in \cite{Bianchi:2025Toller}. 
Fig.~\ref{fig:Contour-Plot} shows concretely, in an example, how the poles in the $\rho$-plane cancel in the sum $t^{(+,\,\rho,k)}_{jlm}(\beta)+t^{(-,\,\rho,k)}_{jlm}(\beta)=d^{(\rho,k)}_{jlm}(\beta)$.

Finally, it is interesting to report the special case $k=j=l=0$ with $\rho\neq 0$, which appears in the Lorentzian Barrett-Crane model \cite{Barrett:1999qw}. In this case, we have the expressions
\begin{equation}
t^{(\pm,\rho,0)}_{000}(\beta) = \frac{\pm 1}{2\,\ii \,\rho} \frac{\ee^{\pm \ii \rho \beta}}{\sinh(\beta)} \, ,\;\; d^{(\rho,0)}_{000}(\beta) = \frac{\sin(\rho\, \beta)}{\rho\, \sinh(\beta)},
\label{eq:Livine-Oriti}
\end{equation}
for the $t$-matrices which match exactly the wedge contributions proposed by Livine and Oriti in \cite{Livine:2002rh, Oriti:2004mu}. As a result, the Livine-Oriti causal version of the Barrett-Crane model is recovered as the limit $\gamma\to \infty$, with fixed areas $\rho=\gamma j$, of the causal vertex \eqref{eq:causal-vertex-def}.

\section{Large-Spin Asymptotics}
\label{sec3:Asymptotics}

For a semiclassical boundary state peaked on a Lorentzian Regge geometry, the large-spin asymptotics ($j_{ab}\gg 1$) of the EPRL model has been studied extensively using both analytical \cite{Barrett:2009gg,Barrett:2009mw,Han:2013hna,Dona:2020xzv,Dona:2020yao,Dona:2022hgr,Han:2020fil} and numerical methods \cite{Speziale:2016axj,Dona:2018nev,Dona:2019dkf,Dona:2020tvv,Gozzini:2021kbt,Dona:2022dxs,Dona:2022yyn}. In this section we first discuss the role of the Regge causal data $s_a=\pm1$ (see Fig.~\ref{fig:Causal-Tetrahedra}) and its application to the EPRL asymptotics. Then, we use the same methods to derive the large-spin asymptotics of the causal vertex introduced in \eqref{eq:causal-vertex-def}.

There are multiple equivalent ways of characterizing the Regge geometry of a Lorentzian $4$-simplex. For instance, one can use length variables, as in the original formulation \cite{Regge:1961px}, or triangle-areas and dihedral angles as done in \cite{Dittrich:2008va,Dona:2017dvf}. Here we adopt variables that make the causal structure manifest: given a Lorentzian $4$-simplex embedded in Minkowski spacetime, we can compute the $4$-normals $N_a^I$ to the tetrahedra $a=1,\ldots,5$ on its boundary. Conversely, a set of five $4$-vectors $N_a^I$ satisfying the closure condition $\sum_a N_a^I=0$ uniquely determines the Regge geometry of a $4$-simplex. This statement provides a generalization of the Minkowski theorem for convex Euclidean polyhedra \cite{Bianchi:2010gc,Minkowski1897} to Lorentzian $4$-simplices (see, e.g., Sec.~3.5.2 of \cite{Wieland:2013ata}). We assume that the boundary tetrahedra are spacelike, and therefore their $4$-normals are timelike, i.e., $\eta_{IJ}N_a^I N_a^J<0$,
where $\eta_{IJ}$ is the Minkowski metric with signature $(-,+,+,+)$. The volume of each tetrahedron is determined by the norm $V_a= \sqrt{-\eta_{IJ}N_a^I N_a^J}\,>0$. The remaining variable is the Regge causal data $s_a$ \cite{Bianchi:2021ric}: we write the $4$-normal as $N_a^I = s_a \, V_a \, \hat{N}_a^I$, where $\hat{N}_a^I$ is a unit timelike vector which we require to be future-pointing with respect to a reference timelike vector $t^I$, i.e., $\eta_{IJ}\, \hat{N}^I_a \, \hat{N}^J_a = -1$ and $\eta_{IJ}\, t^I\, \hat{N}^J_a <0$. The causal variable $s_a=+1\, (-1)$ determines if the $4$-normal $N^I_a$ is future (past) pointing, i.e., $s_a = -\mathrm{sign}(\eta_{IJ}\, t^I\, N^J_a)$, 
(see Fig.~\ref{fig:Causal-Tetrahedra}).

The variables $s_a$, $V_a$, $\hat{N}_a^I$ allow us to describe completely the Regge geometry of a Lorentzian $4$-simplex. While the area $A_{ab}$ of a triangle shared by the tetrahedra $a$ and $b$ is less immediate to write, other quantities have a simple form, such as the boost $\beta_{ab}\geq0$ from the plane of the tetrahedron $a$ to the tetrahedron $b$, which is simply given by $\cosh(\beta_{ab})=-\eta_{IJ}\hat{N}^I_a \hat{N}^J_b$. The Regge action for the Lorentzian $4$-simplex determined by these data is then given by the expression
\begin{equation}
S_{\mathrm{Regge}} = 
\sum_{ab} 
\frac{A_{ab}}{8 \pi G} \, s_a s_b \, \beta_{ab}\, ,
\end{equation}
where the Regge causal data for co-chronal ($s_a s_b=+1$) and anti-chronal ($s_a s_b=-1$) tetrahedra is manifest.

As done in \cite{Barrett:2009gg,Barrett:2009mw,Han:2013hna,Dona:2020xzv,Dona:2020yao,Dona:2022hgr,Han:2020fil}, we consider a semiclassical boundary state $|\Psi_{j_{ab},\zeta_{ab}}\rangle$ peaked on the boundary geometry of a Lorentzian $4$-simplex. The state is labeled by $10$ spins $j_{ab}$ and $20$ spinors $\zeta_{ab}$, and is defined by $5$ coherent intertwiners peaked on the intrinsic geometry of each boundary tetrahedron. The spins and spinors are understood as functions of the $4$-normals $N_a^I$, i.e., $\{ j_{ab}(N_1^{I},\ldots,N_5^{I}),\,\zeta_{ab}(N_1^{I},\ldots,N_5^{I})\}$, and therefore they depend implicitly on the Regge causal data $s_a=\pm 1$. We assume that the boundary state $|\Psi_{j_{ab},\zeta_{ab}}\rangle$ is peaked on a Regge geometry that is Lorentzian and non-degenerate, in the sense that its $4$-volume is non-vanishing, which implies that there are no co-planar tetrahedra, i.e., $\beta_{ab}\neq 0$.

\bigskip

The Wigner $D$-matrix contracted with the spinors $\zeta_{ab}$ and $\mathcal{J}\!\zeta_{ba}$, i.e., $D^{(\rho_{ab},j_{ab})}_{j_{ab} \mathcal{J}\!\zeta_{ba} \, j_{ab} \zeta_{ab}}(g_{ab})$, can be expressed as an integral over an auxiliary spinor $z_{ab}$ with ${\mathbb{C}P^1}$ invariant measure $\dd\Omega_{ab}$, (see Appendix~\ref{sec:app-coherent-Wigner}). In \cite{Barrett:2009gg,Barrett:2009mw,Han:2013hna,Dona:2020xzv,Dona:2020yao,Dona:2022hgr,Han:2020fil}, this formula is used to express the EPRL amplitude of a semiclassical boundary state as the integral
\begin{equation}
\label{eq:EPRL-vertex-coherent} 
\langle A_v^{(\mathrm{EPRL})}|\Psi_{j_{ab},\zeta_{ab}}
\rangle
= \int \prod_{a=2}^{5} \dd{g_a} \int\prod_{ab} \dd{\Omega}_{ab}\,
\ee^{\,\ii\, \mathcal{S}} \, ,
\end{equation}
where the quantity $\mathcal{S}$ playing the role of action in the exponent is linear in $j_{ab}$ and takes the form
\begin{equation}
 \mathcal{S} = \sum_{ab} \qty(\gamma\, j_{ab} \mathcal{B}_{ab} + j_{ab}
\Phi_{ab}) + \ii\, \sum_{ab} j_{ab} Q_{ab}\, ,
\end{equation}
with $\mathcal{B}_{ab}$, $\Phi_{ab}$, $Q_{ab}$ real functions reported in Appendix~\ref{sec:app-coherent-Wigner}. 

The expression \eqref{eq:EPRL-vertex-coherent} allows us to determine the large-spin asymptotics of the EPRL amplitude via a saddle-point analysis of the integral under a uniform rescaling $j_{ab}\to \lambda j_{ab}$ of the spins, with $\lambda \to \infty$, enforcing the condition $j_{ab}\gg 1$. The conditions $\delta \mathcal{S}=0$ and $Q_{ab}=0$ determine two critical points $g_{a}^{(\pm)}$. The function $\mathcal{B}_{ab} = \mathcal{B}(z_{ab},g_b^{-1}g_a)$ at the critical points takes the form
\begin{equation}
    \mathcal{B}_{ab}^{(\pm)} = \pm\, s_a  s_b\, \beta_{ab}\,,
    \label{eq:Bpm}
\end{equation}
where, crucially, the sign $\pm$ is global and does not depend on the wedge $(ab)$. As a result, one obtains the action $\mathcal{S}$ at the critical point in terms of the Regge action:
\begin{equation}
\mathcal{S}^{(\pm)}=\pm \tfrac{1}{\hbar}S_{\mathrm{Regge}}\,+\Upsilon\,,
\end{equation}
where we used the asymptotic expression of the area spectrum $A_{ab}=8\pi G \hbar \gamma \sqrt{j_{ab}(j_{ab}+1)}\sim 8\pi G \hbar\, \gamma j_{ab}$. The phase $\Upsilon=\sum_{ab}j_{ab}(\pm\pi\, \theta(-s_a s_b)-\xi_{ab})$ counts the number of anti-chronal wedges ($s_a s_b=-1$) with the step function $\theta(x)$, and depends on the twist phase $\xi_{ab}$ of the boundary spinors. As a result, the asymptotics is given by a cosine of the Regge action: $\langle A_v^{(\mathrm{EPRL})}|\Psi_{\lambda j_{ab},\zeta_{ab}}\rangle = 
\lambda^{-12} \mathcal{N}^{-1}\,\ee^{\ii \lambda \Upsilon} \, \cos(\lambda S_{\mathrm{Regge}}/\hbar+\mu) + \order{\lambda^{-13}}$, where $\mathcal{N}$ depends on the Hessian $H^{(+)}=\overline{H^{(-)}}$ and $\mu$ is its Maslov index \cite{Dona:2019dkf,Barrett:2009gg,Barrett:2009mw,Dona:2020yao,Dona:2022hgr,Han:2013hna,Han:2020fil,Dona:2020xzv}.

Using the same techniques, we can determine the large-spin asymptotics of the causal vertex \eqref{eq:causal-vertex-def} for a Lorentzian semiclassical boundary state. The causal amplitude is:
\begin{equation}
\label{eq:Causal-vertex-coherent}
\langle 
A_v^{(\sigma_a\sigma_b)}| \Psi_{j_{ab},\zeta_{ab}}\rangle
=\! \int\! \prod_{a=2}^5 \dd{g_a} \, \prod_{ab}  \, T^{(\sigma_a \sigma_b , \gamma j_{ab},j_{ab})}_{j_{ab}\mathcal{J}\!\zeta_{ba}\, j_{ab} \zeta_{ab}} (g_b^{-1} g_a) \, .
\end{equation}
Note that now both the combinatorial causal data $\sigma_a=\pm1$ and the Regge causal data $s_a=\pm 1$ appear: the first in the definition of the vertex (Fig.~\ref{fig:Causal-Spinfoam}), the second in the Regge geometry encoded by the boundary state (Fig.~\ref{fig:Causal-Tetrahedra}).

In \eqref{eq:Causal-vertex-coherent}, $T^{(\sigma_a \sigma_b,\rho,j)}_{j \mathcal{J}\!\xi \, j \zeta}(g)$ denotes the Toller $T$-matrix in the spinorial coherent basis, which can be obtained via the Feynman $\ii \varepsilon$ prescription \eqref{eq:Feynman-ieps} and the expression \eqref{eq:D-coherent-basis} for the Wigner $D$-matrices:
\begin{widetext}
\begin{equation}
\label{eq:Toller-coherent-basis}
T^{(\sigma_a \sigma_b,\,\rho,j)}_{j \mathcal{J}\!\xi\,j\zeta}(g)= 
\tfrac{2j+1}{2\pi\ii} \!\int \!
\tfrac{[z|\dd{z}\rangle\wedge \langle z|\dd{z}]}{\langle z|z\rangle\,\langle g^\dagger z| g^\dagger z\rangle}\,
\Big( \theta\big(\sigma_a \sigma_b\, \mathcal{B}(z,g)\big) \,+\sigma_a \sigma_b\, \delta^{(\rho,j)}\big(\mathcal{B}(z,g)\big)   \Big)\, \ee^{\,\ii\,\rho\, \mathcal{B}(z,g)}
\;\ee^{\,\ii\,j \,\Phi(z,\xi,\zeta,g)}
\;\ee^{-j\,Q(z,\xi,\zeta,g)}\,,
\end{equation}
\end{widetext}
where $\delta^{(\rho,j)}(x)$ is the distribution \eqref{eq:def-delta-rhoj} with support at $x=0$, reported in Appendix \ref{sec:app-coherent-Toller}. In \eqref{eq:Causal-vertex-coherent}, the integral over  the group elements $g_{a}$ and over the  spinors $z_{ab}$ are restricted to a region because of the product over wedges ($\prod_{ab}$) over the Heaviside step functions $\theta(\sigma_a \sigma_b \mathcal{B}_{ab})$ appearing in \eqref{eq:Toller-coherent-basis}. The distribution $\delta^{(\rho,j)}(\mathcal{B}_{ab})$ is supported on the boundary of this region of integration. Now, for non-degenerate Lorentzian boundary data, the two saddle points \eqref{eq:Bpm} can never fall on the boundary of this integration region as $\beta_{ab}\neq 0$. Moreover, two cases arise: either (i) only the saddle point $\mathcal{B}_{ab}^{(+)} = + \, s_a  s_b\, \beta_{ab}$ falls in the region of integration, and therefore the amplitude is oscillatory; or (ii) no saddle point falls in the region of integration and the amplitude is exponentially suppressed. The saddle point corresponding to $\mathcal{B}_{ab}^{(-)} = - \, s_a  s_b\, \beta_{ab}$ cannot contribute to the asymptotics because it never falls in the integration region.

Specifically, for non-degenerate Lorentzian boundary data, we have the large-spin asymptotics:
\begin{align}
\label{eq:causal-rigidity}
\langle 
&A_v^{(\sigma_a\sigma_b)}| \Psi_{\lambda j_{ab},\zeta_{ab}}\rangle
=\\[.5em]
&=\begin{cases}
 \frac{\ee^{\ii (\mu+\lambda \Upsilon)}}{2\,\lambda^{12}\,\mathcal{N}} \, \ee^{+\ii\lambda S_{\mathrm{Regge}}/\hbar} + \order{\lambda^{-13}}\,,  & \sigma_a \sigma_b= +s_a s_b,   \\[1em]
\order{\lambda^{-N}}\,,\;\;\forall N>0\,, &  \sigma_a \sigma_b\neq +s_a s_b,
\end{cases}
\nonumber
\end{align}
where the condition $\sigma_a \sigma_b= +s_a s_b$ is understood as applying to all couples $(ab)$.
This formula shows a form of \emph{causal rigidity}: if we consider the vertex with fixed  causal structure $\sigma_a$ in the class $1\leftrightarrow 4$ (see Fig.~\ref{fig:Causal-Spinfoam}), and vary over semiclassical boundary data encoding a non-degenerate Lorentzian geometry, then the Regge geometries in the causal class $1\leftrightarrow 4$ (see Fig.~\ref{fig:Causal-Tetrahedra}) are weighted by an oscillatory exponential given by the Regge action, while the Regge geometries in the causal class $2\leftrightarrow 3$ are exponentially suppressed. An analogous conclusion applies to the two other causal classes, $2\leftrightarrow 3$ and $0\leftrightarrow 5$. Note that a non-degenerate Lorentzian Regge geometry can never belong to the $0\leftrightarrow 5$ class, because five timelike vectors cannot be all future-pointing, have finite norm, and satisfy the closure constraint $\sum_a N_a^I=0$ .

\section{Discussion}
\label{sec:Discussion}

In this paper, we introduced a causal vertex amplitude for $4$d Lorentzian spinfoams, defined by the expression \eqref{eq:causal-vertex-def}, as briefly summarized here:
\begin{itemize}[leftmargin=1em]
    \item[--] We introduced the new expression \eqref{eq:Feynman-ieps} for  Toller $T$-matrices \cite{Toller:1968gr,Toller:1968pole,Sciarrino:1967}, written in terms of a Feynman $\ii\varepsilon$ prescription applied to the Wigner $D$-matrices. In the literature on harmonic analysis on the Lorentz group \cite{Ruhl:1970lor}, the $T$-matrices are described in terms of their analytic properties (Toller poles). The new formula \eqref{eq:Feynman-ieps} encodes completely the structure of Toller poles (Fig.~\ref{fig:Contour-Plot}) and plays a central role in our construction of the causal vertex amplitude for the EPRL model.

    \item[--] The causal vertex amplitude proposed here replaces the Wigner $D$-matrices, which are associated with wedges in the EPRL model \cite{Engle:2007wy}, with Toller $T$-matrices \eqref{eq:Feynman-ieps}. In this way, the spinfoam vertex amplitude defined in \eqref{eq:causal-vertex-def} depends on the causal structure of the vertex via the orientation $\sigma_a$ of the edges (Fig.~\ref{fig:Causal-Spinfoam}). This construction of the causal vertex for the EPRL model follows the same logic as the construction of the  Livine-Oriti causal vertex \cite{Livine:2002rh,Oriti:2004mu}  for the Barrett-Crane model \cite{Barrett:1999qw}; there, one replaces the function $d^{(\rho,0)}_{000}(\beta)$ with the  functions $t^{(\pm,\rho,0)}_{000}(\beta)$, reported in \eqref{eq:Livine-Oriti}. In fact, as the Barrett-Crane model can be obtained from the EPRL model in the limit $\gamma\to\infty$ with fixed area $\rho=\gamma j$, the Livine-Oriti model is reproduced in the same limit from \eqref{eq:causal-vertex-def}.

    \item[--] We studied the large-spin asymptotics of the causal vertex amplitude for a semiclassical boundary state peaked on a non-degenerate Lorentzian Regge geometry. The integral formula \eqref{eq:Toller-coherent-basis} for the Toller $T$-matrices allows us to employ the same saddle-point techniques developed for the EPRL model \cite{Barrett:2009gg,Barrett:2009mw,Han:2013hna,Dona:2020xzv,Dona:2020yao,Dona:2022hgr,Han:2020fil}. In fact, the formula \eqref{eq:Toller-coherent-basis} provides a realization at the exact level of the \emph{ad hoc} proposal \cite{Bianchi:2021ric} where one inserts in the integral a step function  $\theta(+\sigma_a \sigma_b\, s_a s_b)$. Besides selecting one of the two saddle points, the step function  enforces a form of causal rigidity: it relates the combinatorial causal class ($0\leftrightarrow 5$, $1\leftrightarrow 4$, or $2\leftrightarrow 3$) defined by the edge orientation $\sigma_a$, to the Regge causal  class  ($1\leftrightarrow 4$ or $2\leftrightarrow 3$) defined by the time orientation $s_a$ of the tetrahedra in a non-degenerate Lorentzian $4$-simplex (Fig.~\ref{fig:Spinfoam}). If the two classes agree, then we obtain an exponential of $+\ii$ times the Regge action \eqref{eq:causal-rigidity}; on the other hand, if the two classes do not agree, the vertex amplitude is exponentially suppressed.
\end{itemize}

\noindent Many aspects of the causal vertex amplitude \eqref{eq:causal-vertex-def} remain to be explored. We briefly discuss some of them here:

\begin{enumerate}[leftmargin=2em,label=\roman*.]
    \item Numerical implementations of the EPRL model, such as \texttt{sl2cfoam-next} \cite{Dona:2018nev,Dona:2019dkf,Dona:2020tvv,Gozzini:2021kbt,Dona:2022dxs}, provide a powerful computational tool in spinfoams \cite{Dona:2022yyn}. We expect that the analytic expressions \eqref{eq:t-gamma-simple} and \eqref{eq:APP-t-plus}-\eqref{eq:APP-t-minus} for the Toller $t$-functions will allow us to obtain a numerical implementation of the causal vertex \eqref{eq:causal-vertex-def} analogous to the one 
    based on booster functions for the EPRL model \cite{Speziale:2016axj}. 
    
    \item Despite the unbounded integrals over $SL(2,\mathbb{C})$ in the definition of the EPRL model, once one takes into account its global Lorentz invariance, one finds that the vertex is finite \cite{Engle:2008ev}. The question of finiteness needs to be investigated again for the causal model \eqref{eq:causal-vertex-def}, to check that the Toller poles (Fig.~\ref{fig:Contour-Plot}) do not result in new divergences. We note that, in the Livine-Oriti limit, a naive power-counting of the poles in \eqref{eq:Livine-Oriti} indicates a possible divergence \cite{Pfeiffer:2002ic}, but detailed calculations show that the vertex \cite{Livine:2002rh} is in fact finite \cite{Christensen:2005tr}.  

    \item In our analysis of the large-spin asymptotics for Lorentzian boundary data, we focused on the contribution of the two saddle points $g_a^{(\pm)}$, but we did not investigate the Hessian $H^{(\pm)}$ and its Maslov index at these points. The integral formula \eqref{eq:Toller-coherent-basis} shows that the Hessians $H^{(\pm)}$ for the causal vertex \eqref{eq:causal-vertex-def} coincide with the EPRL ones, for which there is numerical evidence that the relation $H^{(+)}=\overline{H^{(-)}}$ holds \cite{Dona:2020yao}. In fact, the analytical properties relating the Toller $T$-matrices might help prove this relation \cite{Bianchi:2025Toller} and its consequences for the spinfoam propagator \cite{Bianchi:2006uf,Bianchi:2009ri,Bianchi:2011hp}.

    \item In our analysis of the large-spin asymptotics, we focused on non-degenerate Lorentzian boundary data. To complete the analysis of the large-spin asymptotics, it is important to investigate the case of Euclidean and vector geometries, which are  unsuppressed saddle points in the EPRL model \cite{Barrett:2009gg,Barrett:2009mw,Han:2013hna,Dona:2020xzv,Dona:2020yao,Dona:2022hgr,Han:2020fil}, but are not ordinary saddle points here as they are on the boundary of the integration region.

    \item At the level of a single vertex, one can consider a sum over causal structures and define the causal vertex $\langle A_v^{\mathcal{C}_+}|\equiv \sum_{\sigma_a}\langle A_v^{(\sigma_a \sigma_b)}|$. Because of the causal rigidity property \eqref{eq:causal-rigidity}, this causal vertex has asymptotics $\langle A_v^{\mathcal{C}_+}|\Psi_{j_{ab},\zeta_{ab}}\rangle\sim \ee^{+\ii S_{\mathrm{Regge}}/\hbar}$ for non-degenerate Lorentzian boundary states. Alternatively, one can also introduce a \emph{co-causal} vertex defined by $\langle A_v^{\mathcal{C}_-}|\equiv \sum_{\sigma_a}\langle A_v^{(-\sigma_a \sigma_b)}|$, which has asymptotics $\langle A_v^{\mathcal{C}_-}|\Psi_{j_{ab},\zeta_{ab}}\rangle\sim \ee^{-\ii S_{\mathrm{Regge}}/\hbar}$. It would be useful to clarify the relation of the causal vertex $\langle A_v^{\mathcal{C}_+}|$ to Engle's definition of the proper vertex, where a constraint on the $4$-volume orientation results in the selection of a single critical point  \cite{Engle:2011un,Engle:2012yg,Engle:2015mra,Engle:2015zqa}. We note also that, as discussed in \cite{Bianchi:2021ric}, the EPRL model, $\langle A_v^{\mathrm{EPRL}}|=\langle A_v^{\mathcal{C}_+}|+\langle A_v^{\mathcal{C}_-}|+\langle A_v^{\slashed{\mathcal{C}}}|$, also includes the \emph{non-causal} contribution $\langle A_v^{\slashed{\mathcal{C}}}|$ corresponding to the $992$ out of $1024$ configurations $\kappa_{ab}\neq \pm \sigma_a \sigma_b$. It would also be useful to explore the effect of excluding the non-causal contributions, while keeping both causal  terms $\langle A_v^{\mathcal{C}_+}|+\langle A_v^{\mathcal{C}_-}|$.

    \item In this paper we focused on the spinfoam amplitude of a single vertex. This is a building block for constructing the spinfoam path integral with many vertices. At fixed $2$-complex, the edge orientations define a causal set \cite{Bombelli:1987aa,Sorkin:2003bx,Dowker:2005tz,Surya:2019ndm} as discussed in \cite{Bianchi:2021ric,Markopoulou:1997hu,Markopoulou:1999cz,Gupta:1999cp,Livine:2002rh,Pfeiffer:2002ic,Hawkins:2003vc,Oriti:2004mu,Freidel:2005bb,Oriti:2005jr,Oriti:2006wq,Livine:2006xc,Rovelli:2012yy,Bianchi:2012nk,Oriti:2013aqa,Immirzi:2013rka,Cortes:2014oka,Wieland:2014nka,Immirzi:2016nnz,Finocchiaro:2018hks,Jercher:2022mky,Simao:2024don}. An important next step is to explore the various open questions in spinfoams \cite{Engle:2023qsu,Livine:2024hhc}---including the role of bubble divergences, spikes, propagation of curvature, products of cosines and semiclassical limit---at fixed causal orientation of the edges. The new causal rigidity \eqref{eq:causal-rigidity} is expected to tame many of the known issues \cite{Bianchi:2021ric}. It would also be interesting to investigate a sum over both causal and co-causal structures for many vertices, excluding from the path integral the non-causal configurations.
    
    \item Finally, it would be interesting to investigate the effects of the causal spinfoam vertex \eqref{eq:causal-vertex-def} on transition amplitudes of phenomenological relevance, in spinfoam cosmology \cite{Vidotto:2010kw,Bianchi:2010zs,Roken:2010vp,Henderson:2010qd,Bianchi:2011ym,Livine:2011up,Rennert:2013qsa,Rennert:2013pfa,Vilensky:2016tnw,Gozzini:2019nbo,Frisoni:2022urv,Frisoni:2023lvb,Han:2024ydv} and in spinfoam black-to-white hole tunneling \cite{Haggard:2014rza,Christodoulou:2016vny,Christodoulou:2018ryl,Bianchi:2018mml,DAmbrosio:2020mut,Soltani:2021zmv,Christodoulou:2023psv,Frisoni:2023agk,Dona:2024rdq,Rovelli:2024sjl,Han:2024rqb,Dona:2025snr}.

\end{enumerate}

\newpage

\begin{acknowledgments}
We thank Pierre Martin-Dussaud, Pietro Don\`a, Monica Rincon-Ramirez, Hal Haggard, Carlo Rovelli, Jonathan Engle, and Francesca Vidotto for many insightful discussions on the nature of causality in spinfoams. M.G. was partially supported by the \href{https://anid.cl}{Agencia Nacional de Investigación y Desarrollo} (ANID) and \href{https://fulbrightchile.cl/}{Fulbright Chile} through the Fulbright Foreign Student Program and ANID BECAS/Doctorado BIO Fulbright-ANID 56190016.~E.B. is supported by the National Science Foundation, Grants No. PHY-2207851 and PHY-2513194. This work was made possible thanks to the support of the WOST project (\href{https://withoutspacetime.org}{\mbox{withoutspacetime.org}}), funded by the John Templeton Foundation (JTF) under Grant ID 63683. 
\end{acknowledgments}


\onecolumngrid

\appendix


\section{Reduced Toller matrices -- Analytic properties and closed-form expression}
\label{app:FullExpressions}

The Wigner $d$-matrices can be expressed as a sum over hypergeometric functions. Following R\"uhl's phase conventions \cite{Ruhl:1970lor}, we have
\begin{align}
  d^{(\rho,k)}_{jlm}(\beta)
  &= \sqrt{(1+2j)(1+2l)}\,
     \sqrt{\frac{(j-k)!\,(j+k)!\,(-k+l)!\,(k+l)!}
                 {(j-m)!\,(l-m)!\,(j+m)!\,(l+m)!}}
     \;\frac{1}{(1+j+l)!}
     \nonumber\\
  &\quad\times
     \sum_{n_1,n_2} (-1)^{n_1+n_2}
     \binom{j-m}{k-m+n_1}
     \binom{l-m}{k-m+n_2}
     \binom{j+m}{n_1}
     \binom{l+m}{n_2}
     \nonumber\\
  &\quad\times
     (j-k+l+m-n_1-n_2)!\,
     (k-m+n_1+n_2)!\,
     \ee^{\,\beta(-1-k+m-2n_1-\ii\rho)}
     \nonumber\\
  &\quad\times
     {}_{2}F_{1}\Big(
       1+k-m+n_1+n_2,\;
       1+j+\ii\rho;\;
       2+j+l;\;
       1-\ee^{-2\beta}
     \Big)\,,
  \label{eq:d-Ruhl-sum}
\end{align}
where the sums over $n_1,n_2$ run over $n_1 = \max(0,m-k),\ldots,\min(j+m,j-k)$ and $n_2 = \max(0,m-k),\ldots,\min(l+m,l-k)$. With this phase convention, the Wigner $d$-matrices are entire in the complex $\rho$-plane. Under analytic continuation in $\rho$, and assuming $\beta>0$, R\"uhl \cite{Ruhl:1970lor} proves the existence of ``functions of the second kind'' $e^{(\rho,k)}_{jlm}(\beta)$, such that 
$d^{(\rho,k)}_{jlm}(\beta) =e^{(\rho,k)}_{jlm}(\beta)  + (-1)^{j-l}\, e^{(-\rho,-k)}_{ljm}(\beta)  $. We then identify the Toller $t$-matrices as $t^{(+,\,\rho,k)}_{jlm}(\beta) = e^{(\rho,k)}_{jlm}(\beta)$ and $t^{(-,\,\rho,k)}_{jlm}(\beta) =(-1)^{j-l}\, e^{(-\rho,-k)}_{ljm}(\beta) $. 
These functions are defined by the following properties:
\begin{enumerate}
    \item \textbf{Asymptotic properties}:
    
    Rapid decay across the upper (lower) $\rho$-plane for $t^{(+,\rho,k)}_{jlm}(\beta)$ ($t^{(-,\rho,k)}_{jlm}(\beta)$),
    \begin{equation}
    \label{eq:Toller-Property-1}
        t^{(\pm,\rho,k)}_{jlm}(\beta)=\order{\frac{1}{\abs{\rho}^\alpha}} \, , \;\; \textrm{as} \;\abs{\rho}\to \infty\; \textrm{and} \, \begin{cases}
            \alpha \in \mathbb{R}\, &\mathrm{for} \, \Im{\rho}\gtrless 0\\[0.5em]
            0<\alpha<1 &\mathrm{for} \, \Im{\rho}=0
        \end{cases}
    \end{equation}
    \item \textbf{Matching properties}:

    Matching with $d^{(\rho,k)}_{jlm}(\beta)$ across the lower (upper) $\rho$-plane for $t^{(+,\rho,k)}_{jlm}(\beta)$ ($t^{(-,\rho,k)}_{jlm}(\beta)$),
    \begin{equation}
    \label{eq:Toller-Property-2}
        t^{(\pm,\rho,k)}_{jlm}(\beta)= d^{(\rho,k)}_{jlm}(\beta)+\order{\frac{1}{\abs{\rho}^\alpha}} \, , \;\; \textrm{as} \;\abs{\rho}\to \infty\; \textrm{and} \, \begin{cases}
            \alpha \in \mathbb{R}\, &\mathrm{for} \, \Im{\rho}\lessgtr 0\\[0.5em]
            0<\alpha<1 &\mathrm{for} \, \Im{\rho}=0
        \end{cases}
    \end{equation}
    \item \textbf{Pole structure}:

    The Toller $t$-matrices are meromorphic in the $\rho$-plane, with $(j+l+1)$ simple poles on the imaginary axis:
    \begin{equation}
    \label{eq:Toller-Property-3}
    \ii\, \rho =  -j  \,, (-j+1)\, \,, \ldots \, , (l-1) \, ,  l\, .
    \end{equation}
\end{enumerate}

By inspection, and taking into account these properties, we note that the Toller $t$-matrices are related to the Wigner $d$-matrices by the Feynman $\ii \varepsilon$ prescription:
\begin{align}
    t^{(\pm,\,\rho,k)}_{jlm}(\beta)&=\lim_{\varepsilon \to 0^+}\int_{-\infty}^{+\infty}\frac{\dd{\tilde{\rho}}}{2\pi\ii}\;\frac{\pm1}{\tilde{\rho}-\rho\mp\ii\, \varepsilon}\;\qty(\prod_{n=0}^{j+l}\frac{\ii\, \tilde{\rho}- \,(n-j)}{\ii\,\rho-(n-j)} )\;d^{(\tilde{\rho},k)}_{jlm}(\beta)\, \nonumber \\
    &=\lim_{\varepsilon \to 0^+}\int_{-\infty}^{+\infty}\frac{\dd{\tilde{\rho}}}{2\pi\ii}\;\frac{\pm1}{\tilde{\rho}-\rho \mp\ii\, \varepsilon} \; \frac{\Gamma (-j-\ii \rho) \Gamma (l-\ii \tilde{\rho} +1)}{\Gamma (-j-\ii \tilde{\rho} ) \Gamma (l-\ii \rho+1)}\, d^{(\tilde{\rho},k)}_{jlm}(\beta)\,.
\end{align}
To our knowledge, this formula has not appeared before in the literature. An explicit formula of the reduced Toller $t$-matrices can be obtained in terms of  $e^{(\rho,k)}_{jlm}(\beta) $, i.e., explicitly in terms of hypergeometric functions, as 
\begin{align}
\label{eq:APP-t-plus}
	&t_{j l m}^{(+,\rho,k)}(\beta) = \sqrt{(1+2j)(1+2l)}\;
	   \sqrt{\frac{(j-k)!\,(j+k)!\,(l-k)!\,(l+k)!}{(j-m)!\,(j+m)!\,(l-m)!\,(l+m)!}} \nonumber \\
	&\quad \times 
	   \sum_{n_1}
	   \sum_{n_2}
	   \frac{\Gamma(j+k+m-n_1-n_2+i\rho)}{\Gamma(1+j+i\rho)}
	   (-k-m+n_1+n_2)!\;
	   \binom{j-m}{-k-m+n_1}
	   \binom{l-m}{-k-m+n_2}
	   \binom{j+m}{n_1}
	   \binom{l+m}{n_2} \nonumber \\
	&\qquad \times
	   (-1)^{\,j-l+n_1+n_2}\,
	   e^{\,\beta(-1+k+m-2n_2+i\rho)} \;
	   {}_2F_1\!\left(
		 \begin{matrix}
		   1-k-m+n_1+n_2,\; 1+l-i\rho \\
		   1-j-k-m+n_1+n_2-i\rho
		 \end{matrix}
		 ;\, e^{-2\beta}
	   \right)\, ,
\end{align}
with $n_1 = \max(0,m+k),\cdots , \min(j+m,j+k)$ and $n_2 = \max(0,m+k),\cdots , \min(l+m,l+k)$, and
\begin{align}
\label{eq:APP-t-minus}
	&t_{j l m}^{(-,\rho,k)}(\beta) = \sqrt{(1+2l)(1+2j)}\;
	   \sqrt{\frac{(l+k)!\,(l-k)!\,(j+k)!\,(j-k)!}{(l-m)!\,(l+m)!\,(j-m)!\,(j+m)!}} \nonumber \\
	&\quad \times 
	   \sum_{n_1}
	   \sum_{n_2}
	   \frac{\Gamma(l-k+m-n_1-n_2-i\rho)}{\Gamma(1+l-i\rho)}
	   (k-m+n_1+n_2)!\;
	   \binom{l-m}{k-m+n_1}
	   \binom{j-m}{k-m+n_2}
	   \binom{l+m}{n_1}
	   \binom{j+m}{n_2} \nonumber \\
	&\qquad \times
	   (-1)^{n_1+n_2}\,
	   e^{\,\beta(-1-k+m-2n_2-i\rho)} \;
	   {}_2F_1\!\left(
		 \begin{matrix}
		   1+k-m+n_1+n_2,\; 1+j+i\rho \\
		   1-l+k-m+n_1+n_2+i\rho
		 \end{matrix}
		 ;\, e^{-2\beta}
	   \right)\, ,
\end{align}
with $n_1 = \max(0,m-k),\cdots , \min(l+m,l-k)$ and $n_2 = \max(0,m-k),\cdots , \min(j+m,j-k)$.

\newpage

\section{Reduced Toller matrices: Examples}
\label{app:BoosterFunctions}
\vspace{-0.5\baselineskip}

\begin{table*}[h]
\centering
\setlength{\textfloatsep}{5pt plus 2pt minus 2pt}
\renewcommand{\arraystretch}{2.4} 
\setlength{\tabcolsep}{30pt} 

\caption{Analytic expressions of the reduced Wigner and Toller matrices for $k=1$, $j=1$, $l=1$, and $m=0,-1,1$.}
\label{tab:functions_rho1}

\begin{tabular}{l l} 
\hline \hline 

\multicolumn{2}{c}{$\boldsymbol{k=1,\ j=1,\ l=1,\ m=0}$}\\
\hline
$d^{(\rho,1)}_{110}(\beta)$ & 
$\displaystyle -\frac{12 e^{2 \beta}\left(\left(e^{2 \beta}-1\right)\rho \cos(\beta\rho)-\left(e^{2 \beta}+1\right)\sin(\beta\rho)\right)}
{\left(e^{2 \beta}-1\right)^3\left(\rho+\rho^3\right)}$ \\

$t^{(+,\,\rho,1)}_{110}(\beta)$ & 
$\displaystyle -\frac{3 e^{\ii \beta \rho}\,(\rho+\ii \coth(\beta))\,\csch^2(\beta)}
{2\left(\rho+\rho^3\right)}$ \\

$t^{(-,\,\rho,1)}_{110}(\beta)$ & 
$\displaystyle -\frac{3 e^{-\ii \beta \rho}\,(\rho-\ii \coth(\beta))\,\csch^2(\beta)}
{2\left(\rho+\rho^3\right)}$ \\

\hline
\multicolumn{2}{c}{$\boldsymbol{k=1,\ j=1,\ l=1,\ m=-1}$}\\
\hline
$d^{(\rho,1)}_{11-1}(\beta)$ & 
$\displaystyle \frac{12 e^{3 \beta}\left(-\sin(\beta\rho)+e^{\ii \beta\rho}\rho \sinh(\beta)\left(\cosh(\beta)-\ii\rho \sinh(\beta)\right)\right)}
{\left(e^{2 \beta}-1\right)^3\left(\rho+\rho^3\right)}$ \\

$t^{(+,\,\rho,1)}_{11-1}(\beta)$ & 
$\displaystyle \frac{3 e^{\ii \beta \rho}\,\csch^3(\beta)\left(\ii(1+\rho^2)+\rho\left(\sinh(2\beta)-\ii\rho\cosh(2\beta)\right)\right)}
{4\left(\rho+\rho^3\right)}$ \\

$t^{(-,\,\rho,1)}_{11-1}(\beta)$ & 
$\displaystyle -\frac{3 \ii e^{-\ii \beta \rho}\,\csch^3(\beta)}
{4\left(\rho+\rho^3\right)}$ \\

\hline
\multicolumn{2}{c}{$\boldsymbol{k=1,\ j=1,\ l=1,\ m=1}$}\\
\hline
$d^{(\rho,1)}_{111}(\beta)$ & 
$\displaystyle \frac{6 e^{\beta(3-\ii \rho)}\left(\ii\left(e^{2 \ii \beta \rho}-\rho^2-1\right)+\rho\left(\ii\rho\cosh(2\beta)+\sinh(2\beta)\right)\right)}
{\left(e^{2 \beta}-1\right)^3\left(\rho+\rho^3\right)}$ \\

$t^{(+,\,\rho,1)}_{111}(\beta)$ & 
$\displaystyle \frac{3 \ii e^{\ii \beta \rho}\,\csch^3(\beta)}
{4\left(\rho+\rho^3\right)}$ \\

$t^{(-,\,\rho,1)}_{111}(\beta)$ & 
$\displaystyle \frac{3 \csch^3(\beta)\,(\ii\cos(\beta\rho)+\sin(\beta\rho))\left(-\rho^2+\rho^2\cosh(2\beta)-\ii\rho\sinh(2\beta)-1\right)}
{4\left(\rho+\rho^3\right)}$ \\
\hline \hline 
\end{tabular}

\end{table*}
\FloatBarrier

\section{Wigner $D$-matrices in the spinor bases}
\label{sec:app-coherent-Wigner}

We report the expression for the Wigner $D$-matrices in the spinorial bases \cite{Barrett:2009gg,Barrett:2009mw,Han:2013hna,Dona:2020xzv,Dona:2020yao,Dona:2022hgr,Han:2020fil,Speziale:2016axj}, following the conventions of \cite{Dona:2022hgr}. We use the Dirac notation to describe a spinor $\ket{z}\in \mathbb{C}^2$,
\begin{equation}
\ket{z} \equiv \mqty(z_0 \\ z_1) \, , \quad \bra{z} \equiv (\bar{z}_0,\bar{z}_1)\, .
\end{equation}
The spinorial space is equipped with an inner product $\bra{w}\ket{z}=\bar{w}_0z_0+\bar{w}_1z_1$ and an anti-linear map $\mathcal{J}\!:\mathbb{C}^2\to\mathbb{C}^2$, with
\begin{equation}
    |z] \equiv \mathcal{J}\!\ket{z}  = \mqty(-\bar{z}_1 \\ \bar{z}_0)\, ,
\end{equation}
and $[z| = (|z])^\dagger$. With these definitions, the following identities hold:
\begin{equation}
\bra{w}\ket{z}= \overline{\bra{z}\ket{w}} \, , \qquad [z\!\ket{w} = - [w\!\ket{z} = \overline{\bra{w}\!z]}\, ,
\end{equation}
with the antisymmetric bilinear contraction $[z\!\ket{w} \;=\; z_0\,w_1 - z_1\,w_0 $, such that $[z\!\ket{z}=0$. We now introduce the $\mathrm{SU}(2)$ coherent states in the spin-$j$ representation. For any nonzero spinor $\ket{\zeta}\in\CC^2\setminus\{0\}$, define the (normalized) spin-$j$ coherent state as an expansion in the magnetic basis $\ket{(\rho,k);j,m}$ as \cite{Perelomov:1971bd}
\begin{equation}
\ket{j,\zeta}
\;\equiv\;
\sum_{m=-j}^{+j}
\frac{\sqrt{(2j)!}}{\sqrt{(j+m)!\,(j-m)!}}\,
\frac{(\zeta^0)^{j+m}(\zeta^1)^{j-m}}{\braket{\zeta}^{\,j}}\,
\ket{(\rho,k);j,m}\, .
\end{equation}
The unitary principal series representations of $SL(2,\CC)$ are labeled by $(\rho,j)$ with $\rho\in\mathbb{R}$ and $j\in\{0,\tfrac12,1,\dots\}$. Denoting by $D^{(\rho,j)}(g)$ the corresponding representation operator for $g\in SL(2,\CC)$, we define the coherent-basis Wigner matrix element as the matrix element between $\mathrm{SU}(2)$ coherent states in the lowest-spin block,
\begin{equation}
D^{(\rho,j)}_{j \mathcal{J}\!\xi \, j \zeta}(g)
\;\equiv\;
\bra{j,\mathcal{J}\!\xi}D^{(\rho,j)}(g)\ket{j,\zeta}\, .
\end{equation}
This is the analog of the usual Wigner matrix element $D^{(\rho,k)}_{jm\,jn}(g)=\bra{(\rho,k);jm}D^{(\rho,k)}(g)\ket{(\rho,k);jn}$ for $g\in SL(2,\CC)$, with the magnetic indices replaced by the continuous coherent labels $(\xi,\zeta)$. Remarkably, Wigner $D$-matrix in the coherent basis admits an integral representation in terms of an auxiliary, non-unit spinor $z$ \cite{Barrett:2009gg,Barrett:2009mw,Han:2013hna,Dona:2020xzv,Dona:2020yao,Dona:2022hgr,Han:2020fil,Speziale:2016axj}:
\begin{align}
&D^{(\rho,j)}_{j \mathcal{J}\!\xi \, j \zeta}(g) =
\frac{2j+1}{2\pi \ii}\int_{\mathbb{C}P^1} 
\frac{[z|\dd{z}\rangle\wedge \langle z|\dd{z}]}{\langle z|z\rangle\,\langle g^\dagger z| g^\dagger z\rangle}\,
\;\ee^{\,\ii\,\rho\, \mathcal{B}(z,g)}
\;\ee^{\,\ii\,j \,\Phi(z,\xi,\zeta,g)}
\;\ee^{-j\,Q(z,\xi,\zeta,g)}\,, \label{eq:D-coherent-basis}
\end{align}
where the functions $\mathcal{B}$, $\Phi$, $Q$ are given by
\begin{align}
\mathcal{B}(z,g)\;&=\;\log \qty(\frac{\langle g^\dagger z| g^\dagger z\rangle}{\langle z|z\rangle})\;\;\in \;(-\infty,+\infty)\\[.2em]
\Phi(z,\xi,\zeta,g)\;&=\;\arg\qty(\frac{[\xi|z\rangle\langle g^\dagger z|\zeta\rangle}{\langle\zeta|g^\dagger z\rangle\langle z|\xi]} )\in (-\pi,\pi] \\[.2em]
Q(z,\xi,\zeta,g)\;&=\;\log\qty(\frac{\langle z|z\rangle}{[\xi|z\rangle\langle z|\xi]}\frac{\langle g^\dagger z| g^\dagger z\rangle}{\langle \zeta|g^\dagger z\rangle\langle g^\dagger z|\zeta\rangle} ) \in [0,+\infty)
\, ,
\end{align}
with $\log\qty(\frac{x+\ii y}{x-\ii y})=\ii \,\arg \qty(\frac{x+\ii y}{x-\ii y})$ and $\arg (z)\in(-\pi,+\pi]$.
By construction, the integrand is invariant under complex rescalings $z\mapsto \alpha\,z$, hence the integral is well-defined on $\CC P^1$. In addition, $\mathcal{B}$ and $Q$ are real-valued, while $\Phi$ is a phase defined modulo $2\pi$.

\section{Toller $T$-matrices in the spinor bases}
\label{sec:app-coherent-Toller}

Applying the Feynman $\ii \varepsilon$ prescription \eqref{eq:Feynman-ieps} to the Wigner matrix $D^{(\rho,j)}_{j \mathcal{J}\xi j \zeta}$, the only part that depends on $\rho$, namely $\ee^{\ii \rho \mathcal{B}(z,g)}$ can be isolated, so that
\begin{align}
\label{eq:app-distribution-Exp[x]}
T^{(\sigma,\,\rho,\,j)}_{j \mathcal{J}\xi j \zeta}(g)&=\lim_{\varepsilon \to 0^+}\int_{-\infty}^{+\infty}\frac{\dd{\tilde{\rho}}}{2\pi\ii}\;\frac{\sigma}{\tilde{\rho}-\rho-\ii\sigma\varepsilon} \; F_j(\tilde{\rho},\rho) \; D^{(\tilde{\rho},j)}_{j \mathcal{J}\xi j \zeta}(g) \nonumber \\
&= 
\frac{2j+1}{2\pi \ii}\int_{\mathbb{C}P^1} 
\frac{[z|\dd{z}\rangle\wedge \langle z|\dd{z}]}{\langle z|z\rangle\,\langle g^\dagger z| g^\dagger z\rangle}
\;\qty(  \lim_{\varepsilon\to 0^{+}} \int_{-\infty}^{+\infty} \frac{\dd{\tilde{\rho}}}{2\pi\ii}\;\frac{\sigma}{\tilde{\rho}-\rho-\ii\sigma\varepsilon}\,F_j(\tilde{\rho},\rho)\,e^{\ii \, \tilde{\rho}\, \mathcal{B}(z,g)}    ) 
\;\ee^{\,\ii\,j \,\Phi(z,\xi,\zeta,g)}
\;\ee^{-j\,Q(z,\xi,\zeta,g)}\, \nonumber \\
&= \frac{2j+1}{2\pi \ii}\int_{\mathbb{C}P^1} 
\frac{[z|\dd{z}\rangle\wedge \langle z|\dd{z}]}{\langle z|z\rangle\,\langle g^\dagger z| g^\dagger z\rangle}
\; \Theta_{\sigma,\rho,j} [\mathcal{B}(z,g)] \; \ee^{\,\ii\,\rho \, \mathcal{B}(z,g)}
\;\ee^{\,\ii\,j \,\Phi(z,\xi,\zeta,g)}
\;\ee^{-j\,Q(z,\xi,\zeta,g)}\,, 
\end{align}
where the function $\Theta_{(\sigma,\rho,j)} [\mathcal{B}(z,g)]$ is given by
\begin{align}
\Theta_{(\sigma,\rho,j)}[x] &\equiv \lim_{\varepsilon\to 0^{+}} \int_{-\infty}^{+\infty} \frac{\dd{\tilde{\rho}}}{2\pi\ii}\;\frac{\sigma}{\tilde{\rho}-\rho-\ii\sigma\varepsilon}\,F_j(\tilde{\rho},\rho)\,e^{\ii \, (\tilde{\rho}-\rho)\, x} =\lim_{\varepsilon\to 0^{+}} \int_{-\infty}^{+\infty} \frac{\dd{\tilde{q}}}{2\pi\ii}\;\frac{\sigma}{\tilde{q}-\ii\varepsilon}\,F_j(\rho+\sigma \tilde{q},\rho)\,e^{\ii \, \tilde{q}\, \sigma \, x} \nonumber \\[1em]
&= \theta (\sigma x) + \sigma\, \delta^{(\rho,j)}(x)\, .
\end{align}
Here, $\theta(x)$ is the Heaviside step function, and we define the distribution $\delta^{(\rho,j)}(x)$ as
\begin{equation}
\label{eq:def-delta-rhoj}
\delta^{(\rho,j)}(x) \equiv \sum_{n=0}^{2j} \frac{c_{n+1}^{(\rho,j)}}{(n+1)!} (-\ii)^{n+1}\, \dv[n]{}{x} \delta(x) \, ,
\end{equation}
with coefficients given by
\begin{equation}
c_n^{(\rho,j)}=\eval{\dv[n]{}{\tilde{\rho}} F_j(\tilde{\rho},\rho)}_{\tilde{\rho}=\rho}\, =  \eval{\dv[n]{}{\tilde{\rho}} \qty(\frac{\Gamma (-j-\ii \rho) \Gamma (j-\ii \tilde{\rho} +1)}{\Gamma (-j-\ii \tilde{\rho} ) \Gamma (j-\ii \rho+1)})}_{\tilde{\rho}=\rho}\,.
\end{equation}
Note that the function $F_j$ encodes the Toller poles in a real-valued polynomial of order $2j+1$ in $\tilde{q}$ given by
\begin{equation}
F_j(\rho+\sigma \tilde{q},\rho) = 
\prod_{n=0}^{2j}\frac{\ii \, (\rho+\sigma \tilde{q})- (n-j)}{\ii \, \rho -(n-j)} = 1 + \sum_{n=1}^{2j+1} \frac{c_n^{(\rho,j)}}{n!} (\sigma \, \tilde{q})^n\, . 
\end{equation}
In the limit $j\to\infty$ at fixed $\rho=\gamma j$, the coefficients in the expansion reduce to $c_n^{(\gamma j,j)} = \qty[\ii \, \ln(\frac{\gamma-\ii}{\gamma+\ii})]^n + \order{1/j}$.

\twocolumngrid

\vfill

\providecommand{\href}[2]{#2}\begingroup\raggedright\endgroup



\begin{thebibliography}{200}

\bibitem{Rovelli:2014ssa}
C.~Rovelli and F.~Vidotto, \emph{{Covariant Loop Quantum Gravity}}, Cambridge
  University Press (2014),
  \href{https://doi.org/10.1017/cbo9781107706910}{DOI}.

\bibitem{Ashtekar:2021kfp}
A.~Ashtekar and E.~Bianchi, \emph{{A short review of loop quantum gravity}},
  \href{https://doi.org/10.1088/1361-6633/abed91}{\emph{Rept. Prog. Phys.}
  {\bfseries 84} 042001 (2021)}
  [\href{https://arxiv.org/abs/2104.04394}{{\ttfamily arXiv:2104.04394}}].

\bibitem{Perez:2012wv}
A.~Perez, \emph{{The Spin Foam Approach to Quantum Gravity}},
  \href{https://doi.org/10.12942/lrr-2013-3}{\emph{Living Rev. Rel.} {\bfseries
  16} 3 (2013)} [\href{https://arxiv.org/abs/1205.2019}{{\ttfamily
  arXiv:1205.2019}}].

\bibitem{Engle:2023qsu}
J.~Engle and S.~Speziale, \emph{Spin foams: Foundations},  in \emph{Handbook of
  Quantum Gravity}, C.~Bambi, L.~Modesto and I.~Shapiro, eds., pp.~1--40,
  Springer Nature Singapore (2023),
  \href{https://doi.org/10.1007/978-981-19-3079-9_99-1}{DOI}
  [\href{https://arxiv.org/abs/2310.20147}{{\ttfamily arXiv:2310.20147}}].

\bibitem{Livine:2024hhc}
E.R.~Livine, \emph{Spinfoam models for quantum gravity},  in \emph{Encyclopedia
  of Mathematical Physics (Second Edition)}, (Oxford), pp.~507--519, Academic
  Press (2025), \href{https://doi.org/10.1016/B978-0-323-95703-8.00253-6}{DOI}
  [\href{https://arxiv.org/abs/2403.09364}{{\ttfamily arXiv:2403.09364}}].

\bibitem{Engle:2007wy}
J.~Engle, E.~Livine, R.~Pereira and C.~Rovelli, \emph{{LQG vertex with finite
  Immirzi parameter}},
  \href{https://doi.org/10.1016/j.nuclphysb.2008.02.018}{\emph{Nucl. Phys. B}
  {\bfseries 799} 136 (2008)}
  [\href{https://arxiv.org/abs/0711.0146}{{\ttfamily arXiv:0711.0146}}].

\bibitem{Reisenberger:1996pu}
M.P.~Reisenberger and C.~Rovelli, \emph{{'Sum over surfaces' form of loop
  quantum gravity}},
  \href{https://doi.org/10.1103/PhysRevD.56.3490}{\emph{Phys. Rev. D}
  {\bfseries 56} 3490 (1997)}
  [\href{https://arxiv.org/abs/gr-qc/9612035}{{\ttfamily
  arXiv:gr-qc/9612035}}].

\bibitem{Markopoulou:1997wi}
F.~Markopoulou and L.~Smolin, \emph{{Causal evolution of spin networks}},
  \href{https://doi.org/10.1016/S0550-3213(97)00488-4}{\emph{Nucl. Phys. B}
  {\bfseries 508} 409 (1997)}
  [\href{https://arxiv.org/abs/gr-qc/9702025}{{\ttfamily
  arXiv:gr-qc/9702025}}].

\bibitem{Bianchi:2021ric}
E.~Bianchi and P.~Martin-Dussaud, \emph{{Causal Structure in Spin Foams}},
  \href{https://doi.org/10.3390/universe10040181}{\emph{Universe} {\bfseries
  10} 181 (2024)} [\href{https://arxiv.org/abs/2109.00986}{{\ttfamily
  arXiv:2109.00986}}].

\bibitem{Markopoulou:1997hu}
F.~Markopoulou and L.~Smolin, \emph{{Quantum geometry with intrinsic local
  causality}}, \href{https://doi.org/10.1103/PhysRevD.58.084032}{\emph{Phys.
  Rev. D} {\bfseries 58} 084032 (1998)}
  [\href{https://arxiv.org/abs/gr-qc/9712067}{{\ttfamily
  arXiv:gr-qc/9712067}}].

\bibitem{Markopoulou:1999cz}
F.~Markopoulou, \emph{{Quantum causal histories}},
  \href{https://doi.org/10.1088/0264-9381/17/10/302}{\emph{Class. Quant. Grav.}
  {\bfseries 17} 2059 (2000)}
  [\href{https://arxiv.org/abs/hep-th/9904009}{{\ttfamily
  arXiv:hep-th/9904009}}].

\bibitem{Gupta:1999cp}
S.~Gupta, \emph{{Causality in spin foam models}},
  \href{https://doi.org/10.1103/PhysRevD.61.064014}{\emph{Phys. Rev. D}
  {\bfseries 61} 064014 (2000)}
  [\href{https://arxiv.org/abs/gr-qc/9908018}{{\ttfamily
  arXiv:gr-qc/9908018}}].

\bibitem{Livine:2002rh}
E.R.~Livine and D.~Oriti, \emph{{Implementing causality in the spin foam
  quantum geometry}},
  \href{https://doi.org/10.1016/s0550-3213(03)00378-x}{\emph{Nucl. Phys. B}
  {\bfseries 663} 231 (2003)}
  [\href{https://arxiv.org/abs/gr-qc/0210064}{{\ttfamily
  arXiv:gr-qc/0210064}}].

\bibitem{Oriti:2004mu}
D.~Oriti, \emph{{The Feynman propagator for spin foam quantum gravity}},
  \href{https://doi.org/10.1103/PhysRevLett.94.111301}{\emph{Phys. Rev. Lett.}
  {\bfseries 94} 111301 (2005)}
  [\href{https://arxiv.org/abs/gr-qc/0410134}{{\ttfamily
  arXiv:gr-qc/0410134}}].

\bibitem{Pfeiffer:2002ic}
H.~Pfeiffer, \emph{{On the causal Barrett-Crane model: Measure, coupling
  constant, Wick rotation, symmetries and observables}},
  \href{https://doi.org/10.1103/PhysRevD.67.064022}{\emph{Phys. Rev. D}
  {\bfseries 67} 064022 (2003)}
  [\href{https://arxiv.org/abs/gr-qc/0212049}{{\ttfamily
  arXiv:gr-qc/0212049}}].

\bibitem{Hawkins:2003vc}
E.~Hawkins, F.~Markopoulou and H.~Sahlmann, \emph{{Evolution in quantum causal
  histories}}, \href{https://doi.org/10.1088/0264-9381/20/16/320}{\emph{Class.
  Quant. Grav.} {\bfseries 20} 3839 (2003)}
  [\href{https://arxiv.org/abs/hep-th/0302111}{{\ttfamily
  arXiv:hep-th/0302111}}].

\bibitem{Freidel:2005bb}
L.~Freidel and E.R.~Livine, \emph{{Ponzano-Regge model revisited III: Feynman
  diagrams and effective field theory}},
  \href{https://doi.org/10.1088/0264-9381/23/6/012}{\emph{Class. Quant. Grav.}
  {\bfseries 23} 2021 (2006)}
  [\href{https://arxiv.org/abs/hep-th/0502106}{{\ttfamily
  arXiv:hep-th/0502106}}].

\bibitem{Oriti:2005jr}
D.~Oriti, \emph{{Generalised group field theories and quantum gravity
  transition amplitudes}},
  \href{https://doi.org/10.1103/PhysRevD.73.061502}{\emph{Phys. Rev. D}
  {\bfseries 73} 061502 (2006)}
  [\href{https://arxiv.org/abs/gr-qc/0512069}{{\ttfamily
  arXiv:gr-qc/0512069}}].

\bibitem{Oriti:2006wq}
D.~Oriti and T.~Tlas, \emph{{Causality and matter propagation in 3-D spin foam
  quantum gravity}},
  \href{https://doi.org/10.1103/PhysRevD.74.104021}{\emph{Phys. Rev. D}
  {\bfseries 74} 104021 (2006)}
  [\href{https://arxiv.org/abs/gr-qc/0608116}{{\ttfamily
  arXiv:gr-qc/0608116}}].

\bibitem{Livine:2006xc}
E.R.~Livine and D.R.~Terno, \emph{{Quantum causal histories in the light of
  quantum information}},
  \href{https://doi.org/10.1103/PhysRevD.75.084001}{\emph{Phys. Rev. D}
  {\bfseries 75} 084001 (2007)}
  [\href{https://arxiv.org/abs/gr-qc/0611135}{{\ttfamily
  arXiv:gr-qc/0611135}}].

\bibitem{Rovelli:2012yy}
C.~Rovelli and E.~Wilson-Ewing, \emph{{Discrete Symmetries in Covariant LQG}},
  \href{https://doi.org/10.1103/PhysRevD.86.064002}{\emph{Phys. Rev. D}
  {\bfseries 86} 064002 (2012)}
  [\href{https://arxiv.org/abs/1205.0733}{{\ttfamily arXiv:1205.0733}}].

\bibitem{Bianchi:2012nk}
E.~Bianchi and F.~Hellmann, \emph{{The Construction of Spin Foam Vertex
  Amplitudes}}, \href{https://doi.org/10.3842/SIGMA.2013.008}{\emph{SIGMA}
  {\bfseries 9} 008 (2013)} [\href{https://arxiv.org/abs/1207.4596}{{\ttfamily
  arXiv:1207.4596}}].

\bibitem{Oriti:2013aqa}
D.~Oriti, \emph{{Group field theory as the 2nd quantization of Loop Quantum
  Gravity}}, \href{https://doi.org/10.1088/0264-9381/33/8/085005}{\emph{Class.
  Quant. Grav.} {\bfseries 33} 085005 (2016)}
  [\href{https://arxiv.org/abs/1310.7786}{{\ttfamily arXiv:1310.7786}}].

\bibitem{Immirzi:2013rka}
G.~Immirzi, \emph{{A note on the spinor construction of Spin Foam amplitudes}},
  \href{https://doi.org/10.1088/0264-9381/31/9/095016}{\emph{Class. Quant.
  Grav.} {\bfseries 31} 095016 (2014)}
  [\href{https://arxiv.org/abs/1311.6942}{{\ttfamily arXiv:1311.6942}}].

\bibitem{Cortes:2014oka}
M.~Cort\^es and L.~Smolin, \emph{{Spin foam models as energetic causal sets}},
  \href{https://doi.org/10.1103/PhysRevD.93.084039}{\emph{Phys. Rev. D}
  {\bfseries 93} 084039 (2016)}
  [\href{https://arxiv.org/abs/1407.0032}{{\ttfamily arXiv:1407.0032}}].

\bibitem{Wieland:2014nka}
W.M.~Wieland, \emph{{A new action for simplicial gravity in four dimensions}},
  \href{https://doi.org/10.1088/0264-9381/32/1/015016}{\emph{Class. Quant.
  Grav.} {\bfseries 32} 015016 (2015)}
  [\href{https://arxiv.org/abs/1407.0025}{{\ttfamily arXiv:1407.0025}}].

\bibitem{Immirzi:2016nnz}
G.~Immirzi, \emph{{Causal spin foams}},
  \href{https://arxiv.org/abs/1610.04462}{{\ttfamily arXiv:1610.04462}}.

\bibitem{Finocchiaro:2018hks}
M.~Finocchiaro and D.~Oriti, \emph{{Spin foam models and the Duflo map}},
  \href{https://doi.org/10.1088/1361-6382/ab58da}{\emph{Class. Quant. Grav.}
  {\bfseries 37} 015010 (2020)}
  [\href{https://arxiv.org/abs/1812.03550}{{\ttfamily arXiv:1812.03550}}].

\bibitem{Jercher:2022mky}
A.F.~Jercher, D.~Oriti and A.G.A.~Pithis, \emph{{Complete Barrett-Crane model
  and its causal structure}},
  \href{https://doi.org/10.1103/PhysRevD.106.066019}{\emph{Phys. Rev. D}
  {\bfseries 106} 066019 (2022)}
  [\href{https://arxiv.org/abs/2206.15442}{{\ttfamily arXiv:2206.15442}}].

\bibitem{Simao:2024don}
J.D.~Sim{\~a}o, \emph{{A new 2+1 coherent spin-foam vertex for quantum
  gravity}}, \href{https://doi.org/10.1088/1361-6382/ad721e}{\emph{Class.
  Quant. Grav.} {\bfseries 41} 195015 (2024)}
  [\href{https://arxiv.org/abs/2402.05993}{{\ttfamily arXiv:2402.05993}}].

\bibitem{Oriti:2025uad}
D.~Oriti, \emph{{Quantum information elements in Quantum Gravity states and
  processes}},  \href{https://arxiv.org/abs/2502.21234}{{\ttfamily
  arXiv:2502.21234}}.

\bibitem{Bombelli:1987aa}
L.~Bombelli, J.~Lee, D.~Meyer and R.~Sorkin, \emph{{Space-Time as a Causal
  Set}}, \href{https://doi.org/10.1103/PhysRevLett.59.521}{\emph{Phys. Rev.
  Lett.} {\bfseries 59} 521 (1987)}.

\bibitem{Sorkin:2003bx}
R.D.~Sorkin, \emph{{Causal sets: Discrete gravity}},  in \emph{{School on
  Quantum Gravity}}, pp.~305--327, 9, 2003,
  \href{https://doi.org/10.1007/0-387-24992-3_7}{DOI}
  [\href{https://arxiv.org/abs/gr-qc/0309009}{{\ttfamily
  arXiv:gr-qc/0309009}}].

\bibitem{Dowker:2005tz}
F.~Dowker, \emph{{Causal sets and the deep structure of spacetime}},  in
  \emph{{100 Years Of Relativity}: {space-time structure: Einstein and
  beyond}}, A.~Ashtekar, ed., pp.~445--464 (2005),
  \href{https://doi.org/10.1142/9789812700988_0016}{DOI}
  [\href{https://arxiv.org/abs/gr-qc/0508109}{{\ttfamily
  arXiv:gr-qc/0508109}}].

\bibitem{Surya:2019ndm}
S.~Surya, \emph{{The causal set approach to quantum gravity}},
  \href{https://doi.org/10.1007/s41114-019-0023-1}{\emph{Living Rev. Rel.}
  {\bfseries 22} 5 (2019)} [\href{https://arxiv.org/abs/1903.11544}{{\ttfamily
  arXiv:1903.11544}}].

\bibitem{Teitelboim:1981ua}
C.~Teitelboim, \emph{{Quantum Mechanics of the Gravitational Field}},
  \href{https://doi.org/10.1103/PhysRevD.25.3159}{\emph{Phys. Rev. D}
  {\bfseries 25} 3159 (1982)}.

\bibitem{Barrett:1999qw}
J.W.~Barrett and L.~Crane, \emph{{A Lorentzian signature model for quantum
  general relativity}},
  \href{https://doi.org/10.1088/0264-9381/17/16/302}{\emph{Class. Quant. Grav.}
  {\bfseries 17} 3101 (2000)}
  [\href{https://arxiv.org/abs/gr-qc/9904025}{{\ttfamily
  arXiv:gr-qc/9904025}}].

\bibitem{Ruhl:1970lor}
W.~R\"uhl, \emph{The Lorentz Group and Harmonic Analysis}, Mathematical physics
  monograph series, W. A. Benjamin (1970).

\bibitem{Toller:1968gr}
M.~Toller, \emph{On the group-theoretical approach to complex angular momentum
  and signature}, \href{https://doi.org/10.1007/BF02743789}{\emph{Nuovo Cim. A}
  {\bfseries 54} 295 (1968)}.

\bibitem{Toller:1968pole}
M.~Toller, \emph{{An expansion of the scattering amplitude at vanishing
  four-momentum transfer using the representations of the Lorentz group}},
  \href{https://doi.org/10.1007/BF02721717}{\emph{Nuovo Cim. A} {\bfseries 53}
  671 (1968)}.

\bibitem{Sciarrino:1967}
A.~Sciarrino and M.~Toller, \emph{Decomposition of the unitary irreducible
  representations of the group {SL(2C)} restricted to the subgroup {SU(1, 1)}},
  \href{https://doi.org/10.1063/1.1705341}{\emph{Journal of Mathematical
  Physics} {\bfseries 8} 1252 (1967)}.

\bibitem{Taylor:1967xsb}
J.C.~Taylor, \emph{{Factorization of Regge and Toller poles}},
  \href{https://doi.org/10.1016/0550-3213(67)90056-9}{\emph{Nucl. Phys. B}
  {\bfseries 3} 504 (1967)}.

\bibitem{Jones:1969cd}
H.F.~Jones, \emph{{Covariant O(4) propagators and Toller poles}},
  \href{https://doi.org/10.1007/BF02756347}{\emph{Nuovo Cim. A} {\bfseries 59}
  81 (1969)}.

\bibitem{Bianchi:2025Toller}
E.~Bianchi, C.~Chen and M.~Gamonal, ``Toller functions and the {F}eynman
  {$\mathrm{i} \varepsilon$} prescription in spinfoams.'' To appear (2026).

\bibitem{Regge:1961px}
T.~Regge, \emph{{General Relativity Without Coordinates}},
  \href{https://doi.org/10.1007/bf02733251}{\emph{Nuovo Cim.} {\bfseries 19}
  558 (1961)}.

\bibitem{Barrett:2009gg}
J.W.~Barrett, R.J.~Dowdall, W.J.~Fairbairn, H.~Gomes and F.~Hellmann,
  \emph{{Asymptotic analysis of the EPRL four-simplex amplitude}},
  \href{https://doi.org/10.1063/1.3244218}{\emph{J. Math. Phys.} {\bfseries 50}
  112504 (2009)} [\href{https://arxiv.org/abs/0902.1170}{{\ttfamily
  arXiv:0902.1170}}].

\bibitem{Barrett:2009mw}
J.W.~Barrett, R.J.~Dowdall, W.J.~Fairbairn, F.~Hellmann and R.~Pereira,
  \emph{{Lorentzian spin foam amplitudes: Graphical calculus and asymptotics}},
  \href{https://doi.org/10.1088/0264-9381/27/16/165009}{\emph{Class. Quant.
  Grav.} {\bfseries 27} 165009 (2010)}
  [\href{https://arxiv.org/abs/0907.2440}{{\ttfamily arXiv:0907.2440}}].

\bibitem{Han:2013hna}
M.~Han, \emph{{On Spinfoam Models in Large Spin Regime}},
  \href{https://doi.org/10.1088/0264-9381/31/1/015004}{\emph{Class. Quant.
  Grav.} {\bfseries 31} 015004 (2014)}
  [\href{https://arxiv.org/abs/1304.5627}{{\ttfamily arXiv:1304.5627}}].

\bibitem{Dona:2020xzv}
P.~Dona, M.~Fanizza, P.~Martin-Dussaud and S.~Speziale, \emph{{Asymptotics of
  $\mathrm {SL}(2,{{\mathbb {C}}})$ coherent invariant tensors}},
  \href{https://doi.org/10.1007/s00220-021-04154-3}{\emph{Commun. Math. Phys.}
  {\bfseries 389} 399 (2022)}
  [\href{https://arxiv.org/abs/2011.13909}{{\ttfamily arXiv:2011.13909}}].

\bibitem{Dona:2020yao}
P.~Dona and S.~Speziale, \emph{{Asymptotics of lowest unitary SL(2,C)
  invariants on graphs}},
  \href{https://doi.org/10.1103/PhysRevD.102.086016}{\emph{Phys. Rev. D}
  {\bfseries 102} 086016 (2020)}
  [\href{https://arxiv.org/abs/2007.09089}{{\ttfamily arXiv:2007.09089}}].

\bibitem{Dona:2022hgr}
P.~Dona, \emph{{Geometry from local flatness in Lorentzian spin foam
  theories}}, \href{https://doi.org/10.1103/PhysRevD.107.066011}{\emph{Phys.
  Rev. D} {\bfseries 107} 066011 (2023)}
  [\href{https://arxiv.org/abs/2211.04743}{{\ttfamily arXiv:2211.04743}}].

\bibitem{Han:2020fil}
M.~Han, Z.~Huang, H.~Liu and D.~Qu, \emph{{Numerical computations of
  next-to-leading order corrections in spinfoam large-$j$ asymptotics}},
  \href{https://doi.org/10.1103/PhysRevD.102.124010}{\emph{Phys. Rev. D}
  {\bfseries 102} 124010 (2020)}
  [\href{https://arxiv.org/abs/2007.01998}{{\ttfamily arXiv:2007.01998}}].

\bibitem{Speziale:2016axj}
S.~Speziale, \emph{{Boosting Wigner{\textquoteright}s nj-symbols}},
  \href{https://doi.org/10.1063/1.4977752}{\emph{J. Math. Phys.} {\bfseries 58}
  032501 (2017)} [\href{https://arxiv.org/abs/1609.01632}{{\ttfamily
  arXiv:1609.01632}}].

\bibitem{Dona:2018nev}
P.~Dona and G.~Sarno, \emph{{Numerical methods for EPRL spin foam transition
  amplitudes and Lorentzian recoupling theory}},
  \href{https://doi.org/10.1007/s10714-018-2452-7}{\emph{Gen. Rel. Grav.}
  {\bfseries 50} 127 (2018)}
  [\href{https://arxiv.org/abs/1807.03066}{{\ttfamily arXiv:1807.03066}}].

\bibitem{Dona:2019dkf}
P.~Don\`a, M.~Fanizza, G.~Sarno and S.~Speziale, \emph{{Numerical study of the
  Lorentzian Engle-Pereira-Rovelli-Livine spin foam amplitude}},
  \href{https://doi.org/10.1103/PhysRevD.100.106003}{\emph{Phys. Rev. D}
  {\bfseries 100} 106003 (2019)}
  [\href{https://arxiv.org/abs/1903.12624}{{\ttfamily arXiv:1903.12624}}].

\bibitem{Dona:2020tvv}
P.~Don\`a, F.~Gozzini and G.~Sarno, \emph{{Numerical analysis of spin foam
  dynamics and the flatness problem}},
  \href{https://doi.org/10.1103/PhysRevD.102.106003}{\emph{Phys. Rev. D}
  {\bfseries 102} 106003 (2020)}
  [\href{https://arxiv.org/abs/2004.12911}{{\ttfamily arXiv:2004.12911}}].

\bibitem{Gozzini:2021kbt}
F.~Gozzini, \emph{{A high-performance code for EPRL spin foam amplitudes}},
  \href{https://doi.org/10.1088/1361-6382/ac2b0b}{\emph{Class. Quant. Grav.}
  {\bfseries 38} 225010 (2021)}
  [\href{https://arxiv.org/abs/2107.13952}{{\ttfamily arXiv:2107.13952}}].

\bibitem{Dona:2022dxs}
P.~Dona and P.~Frisoni, \emph{{How-to Compute EPRL Spin Foam Amplitudes}},
  \href{https://doi.org/10.3390/universe8040208}{\emph{Universe} {\bfseries 8}
  208 (2022)} [\href{https://arxiv.org/abs/2202.04360}{{\ttfamily
  arXiv:2202.04360}}].

\bibitem{Dona:2022yyn}
P.~Dona, M.~Han and H.~Liu, \emph{{Spinfoams and High-Performance Computing}},
  in \emph{{Handbook of Quantum Gravity}}, C.~Bambi, L.~Modesto and I.~Shapiro,
  eds., pp.~1--38 (2023),
  \href{https://doi.org/10.1007/978-981-19-3079-9_100-1}{DOI}
  [\href{https://arxiv.org/abs/2212.14396}{{\ttfamily arXiv:2212.14396}}].

\bibitem{Engle:2011un}
J.~Engle, \emph{{Proposed proper Engle-Pereira-Rovelli-Livine vertex
  amplitude}}, \href{https://doi.org/10.1103/PhysRevD.87.084048}{\emph{Phys.
  Rev. D} {\bfseries 87} 084048 (2013)}
  [\href{https://arxiv.org/abs/1111.2865}{{\ttfamily arXiv:1111.2865}}].

\bibitem{Engle:2012yg}
J.~Engle, \emph{{A spin-foam vertex amplitude with the correct semiclassical
  limit}}, \href{https://doi.org/10.1016/j.physletb.2013.06.024}{\emph{Phys.
  Lett. B} {\bfseries 724} 333 (2013)}
  [\href{https://arxiv.org/abs/1201.2187}{{\ttfamily arXiv:1201.2187}}].

\bibitem{Engle:2015mra}
J.~Engle and A.~Zipfel, \emph{{Lorentzian proper vertex amplitude: Classical
  analysis and quantum derivation}},
  \href{https://doi.org/10.1103/PhysRevD.94.064024}{\emph{Phys. Rev. D}
  {\bfseries 94} 064024 (2016)}
  [\href{https://arxiv.org/abs/1502.04640}{{\ttfamily arXiv:1502.04640}}].

\bibitem{Engle:2015zqa}
J.~Engle, I.~Vilensky and A.~Zipfel, \emph{{Lorentzian proper vertex amplitude:
  Asymptotics}}, \href{https://doi.org/10.1103/PhysRevD.94.064025}{\emph{Phys.
  Rev. D} {\bfseries 94} 064025 (2016)}
  [\href{https://arxiv.org/abs/1505.06683}{{\ttfamily arXiv:1505.06683}}].

\bibitem{Engle:2008ev}
J.~Engle and R.~Pereira, \emph{{Regularization and finiteness of the Lorentzian
  LQG vertices}}, \href{https://doi.org/10.1103/PhysRevD.79.084034}{\emph{Phys.
  Rev. D} {\bfseries 79} 084034 (2009)}
  [\href{https://arxiv.org/abs/0805.4696}{{\ttfamily arXiv:0805.4696}}].

\bibitem{Dona:2021ldn}
P.~Don{\`a}, F.~Gozzini and A.~Nicotra, \emph{{Wick rotation for spin foam
  quantum gravity}},
  \href{https://doi.org/10.1103/PhysRevD.104.126008}{\emph{Phys. Rev. D}
  {\bfseries 104} 126008 (2021)}
  [\href{https://arxiv.org/abs/2106.14672}{{\ttfamily arXiv:2106.14672}}].

\bibitem{Dittrich:2008va}
B.~Dittrich and S.~Speziale, \emph{{Area-angle variables for general
  relativity}}, \href{https://doi.org/10.1088/1367-2630/10/8/083006}{\emph{New
  J. Phys.} {\bfseries 10} 083006 (2008)}
  [\href{https://arxiv.org/abs/0802.0864}{{\ttfamily arXiv:0802.0864}}].

\bibitem{Dona:2017dvf}
P.~Don{\`a}, M.~Fanizza, G.~Sarno and S.~Speziale, \emph{{SU(2) graph
  invariants, Regge actions and polytopes}},
  \href{https://doi.org/10.1088/1361-6382/aaa53a}{\emph{Class. Quant. Grav.}
  {\bfseries 35} 045011 (2018)}
  [\href{https://arxiv.org/abs/1708.01727}{{\ttfamily arXiv:1708.01727}}].

\bibitem{Bianchi:2010gc}
E.~Bianchi, P.~Dona and S.~Speziale, \emph{{Polyhedra in loop quantum
  gravity}}, \href{https://doi.org/10.1103/PhysRevD.83.044035}{\emph{Phys. Rev.
  D} {\bfseries 83} 044035 (2011)}
  [\href{https://arxiv.org/abs/1009.3402}{{\ttfamily arXiv:1009.3402}}].

\bibitem{Minkowski1897}
H.~Minkowski, \emph{Allgemeine lehrs{\"a}tze {\"u}ber die convexen polyeder},
  {\emph{Nachr. Ges. Wiss., G{\"o}ttingen} 198 (1897)}.

\bibitem{Wieland:2013ata}
W.M.~Wieland, \emph{{The Chiral Structure of Loop Quantum Gravity}}, Ph.D.
  thesis, Aix-Marseille U., 2013.

\bibitem{Christensen:2005tr}
J.D.~Christensen, \emph{{Finiteness of Lorentzian 10J symbols and partition
  functions}}, \href{https://doi.org/10.1088/0264-9381/23/5/013}{\emph{Class.
  Quant. Grav.} {\bfseries 23} 1679 (2006)}
  [\href{https://arxiv.org/abs/gr-qc/0512004}{{\ttfamily
  arXiv:gr-qc/0512004}}].

\bibitem{Bianchi:2006uf}
E.~Bianchi, L.~Modesto, C.~Rovelli and S.~Speziale, \emph{{Graviton propagator
  in loop quantum gravity}},
  \href{https://doi.org/10.1088/0264-9381/23/23/024}{\emph{Class. Quant. Grav.}
  {\bfseries 23} 6989 (2006)}
  [\href{https://arxiv.org/abs/gr-qc/0604044}{{\ttfamily
  arXiv:gr-qc/0604044}}].

\bibitem{Bianchi:2009ri}
E.~Bianchi, E.~Magliaro and C.~Perini, \emph{{LQG propagator from the new spin
  foams}}, \href{https://doi.org/10.1016/j.nuclphysb.2009.07.016}{\emph{Nucl.
  Phys. B} {\bfseries 822} 245 (2009)}
  [\href{https://arxiv.org/abs/0905.4082}{{\ttfamily arXiv:0905.4082}}].

\bibitem{Bianchi:2011hp}
E.~Bianchi and Y.~Ding, \emph{{Lorentzian spinfoam propagator}},
  \href{https://doi.org/10.1103/PhysRevD.86.104040}{\emph{Phys. Rev. D}
  {\bfseries 86} 104040 (2012)}
  [\href{https://arxiv.org/abs/1109.6538}{{\ttfamily arXiv:1109.6538}}].

\bibitem{Vidotto:2010kw}
F.~Vidotto, \emph{{Spinfoam Cosmology: quantum cosmology from the full
  theory}}, \href{https://doi.org/10.1088/1742-6596/314/1/012049}{\emph{J.
  Phys. Conf. Ser.} {\bfseries 314} 012049 (2011)}
  [\href{https://arxiv.org/abs/1011.4705}{{\ttfamily arXiv:1011.4705}}].

\bibitem{Bianchi:2010zs}
E.~Bianchi, C.~Rovelli and F.~Vidotto, \emph{{Towards Spinfoam Cosmology}},
  \href{https://doi.org/10.1103/PhysRevD.82.084035}{\emph{Phys. Rev. D}
  {\bfseries 82} 084035 (2010)}
  [\href{https://arxiv.org/abs/1003.3483}{{\ttfamily arXiv:1003.3483}}].

\bibitem{Roken:2010vp}
C.~Roken, \emph{{First-order quantum-gravitational correction from covariant,
  holomorphic spinfoam cosmology}},
  \href{https://doi.org/10.1142/S0218271813500053}{\emph{Int. J. Mod. Phys. D}
  {\bfseries 22} 1350005 (2015)}
  [\href{https://arxiv.org/abs/1011.3335}{{\ttfamily arXiv:1011.3335}}].

\bibitem{Henderson:2010qd}
A.~Henderson, C.~Rovelli, F.~Vidotto and E.~Wilson-Ewing, \emph{{Local spinfoam
  expansion in loop quantum cosmology}},
  \href{https://doi.org/10.1088/0264-9381/28/2/025003}{\emph{Class. Quant.
  Grav.} {\bfseries 28} 025003 (2011)}
  [\href{https://arxiv.org/abs/1010.0502}{{\ttfamily arXiv:1010.0502}}].

\bibitem{Bianchi:2011ym}
E.~Bianchi, T.~Krajewski, C.~Rovelli and F.~Vidotto, \emph{{Cosmological
  constant in spinfoam cosmology}},
  \href{https://doi.org/10.1103/PhysRevD.83.104015}{\emph{Phys. Rev. D}
  {\bfseries 83} 104015 (2011)}
  [\href{https://arxiv.org/abs/1101.4049}{{\ttfamily arXiv:1101.4049}}].

\bibitem{Livine:2011up}
E.R.~Livine and M.~Martin-Benito, \emph{{Classical Setting and Effective
  Dynamics for Spinfoam Cosmology}},
  \href{https://doi.org/10.1088/0264-9381/30/3/035006}{\emph{Class. Quant.
  Grav.} {\bfseries 30} 035006 (2013)}
  [\href{https://arxiv.org/abs/1111.2867}{{\ttfamily arXiv:1111.2867}}].

\bibitem{Rennert:2013qsa}
J.~Rennert and D.~Sloan, \emph{{Anisotropic Spinfoam Cosmology}},
  \href{https://doi.org/10.1088/0264-9381/31/1/015017}{\emph{Class. Quant.
  Grav.} {\bfseries 31} 015017 (2014)}
  [\href{https://arxiv.org/abs/1308.0687}{{\ttfamily arXiv:1308.0687}}].

\bibitem{Rennert:2013pfa}
J.~Rennert and D.~Sloan, \emph{{A Homogeneous Model of Spinfoam Cosmology}},
  \href{https://doi.org/10.1088/0264-9381/30/23/235019}{\emph{Class. Quant.
  Grav.} {\bfseries 30} 235019 (2013)}
  [\href{https://arxiv.org/abs/1304.6688}{{\ttfamily arXiv:1304.6688}}].

\bibitem{Vilensky:2016tnw}
I.~Vilensky, \emph{{Spinfoam cosmology with the proper vertex amplitude}},
  \href{https://doi.org/10.1088/1361-6382/aa91f4}{\emph{Class. Quant. Grav.}
  {\bfseries 34} 225015 (2017)}
  [\href{https://arxiv.org/abs/1611.01508}{{\ttfamily arXiv:1611.01508}}].

\bibitem{Gozzini:2019nbo}
F.~Gozzini and F.~Vidotto, \emph{{Primordial Fluctuations From Quantum
  Gravity}}, \href{https://doi.org/10.3389/fspas.2020.629466}{\emph{Front.
  Astron. Astrophys. Cosmol.} {\bfseries 7} 629466 (2021)}
  [\href{https://arxiv.org/abs/1906.02211}{{\ttfamily arXiv:1906.02211}}].

\bibitem{Frisoni:2022urv}
P.~Frisoni, F.~Gozzini and F.~Vidotto, \emph{{Markov chain Monte Carlo methods
  for graph refinement in spinfoam cosmology}},
  \href{https://doi.org/10.1088/1361-6382/acc5d6}{\emph{Class. Quant. Grav.}
  {\bfseries 40} 105001 (2023)}
  [\href{https://arxiv.org/abs/2207.02881}{{\ttfamily arXiv:2207.02881}}].

\bibitem{Frisoni:2023lvb}
P.~Frisoni, F.~Gozzini and F.~Vidotto, \emph{{Primordial fluctuations from
  quantum gravity: 16-cell topological model}},
  \href{https://arxiv.org/abs/2312.02399}{{\ttfamily arXiv:2312.02399}}.

\bibitem{Han:2024ydv}
M.~Han, H.~Liu, D.~Qu, F.~Vidotto and C.~Zhang, \emph{{Cosmological dynamics
  from covariant loop quantum gravity with scalar matter}},
  \href{https://doi.org/10.1103/PhysRevD.111.086012}{\emph{Phys. Rev. D}
  {\bfseries 111} 086012 (2025)}
  [\href{https://arxiv.org/abs/2402.07984}{{\ttfamily arXiv:2402.07984}}].

\bibitem{Haggard:2014rza}
H.M.~Haggard and C.~Rovelli, \emph{{Quantum-gravity effects outside the horizon
  spark black to white hole tunneling}},
  \href{https://doi.org/10.1103/PhysRevD.92.104020}{\emph{Phys. Rev. D}
  {\bfseries 92} 104020 (2015)}
  [\href{https://arxiv.org/abs/1407.0989}{{\ttfamily arXiv:1407.0989}}].

\bibitem{Christodoulou:2016vny}
M.~Christodoulou, C.~Rovelli, S.~Speziale and I.~Vilensky, \emph{{Planck star
  tunneling time: An astrophysically relevant observable from background-free
  quantum gravity}},
  \href{https://doi.org/10.1103/PhysRevD.94.084035}{\emph{Phys. Rev. D}
  {\bfseries 94} 084035 (2016)}
  [\href{https://arxiv.org/abs/1605.05268}{{\ttfamily arXiv:1605.05268}}].

\bibitem{Christodoulou:2018ryl}
M.~Christodoulou and F.~D'Ambrosio, \emph{{Characteristic time scales for the
  geometry transition of a black hole to a white hole from spinfoams}},
  \href{https://doi.org/10.1088/1361-6382/ad6059}{\emph{Class. Quant. Grav.}
  {\bfseries 41} 195030 (2024)}
  [\href{https://arxiv.org/abs/1801.03027}{{\ttfamily arXiv:1801.03027}}].

\bibitem{Bianchi:2018mml}
E.~Bianchi, M.~Christodoulou, F.~D'Ambrosio, H.M.~Haggard and C.~Rovelli,
  \emph{{White Holes as Remnants: A Surprising Scenario for the End of a Black
  Hole}}, \href{https://doi.org/10.1088/1361-6382/aae550}{\emph{Class. Quant.
  Grav.} {\bfseries 35} 225003 (2018)}
  [\href{https://arxiv.org/abs/1802.04264}{{\ttfamily arXiv:1802.04264}}].

\bibitem{DAmbrosio:2020mut}
F.~D'Ambrosio, M.~Christodoulou, P.~Martin-Dussaud, C.~Rovelli and F.~Soltani,
  \emph{{End of a black hole{\textquoteright}s evaporation}},
  \href{https://doi.org/10.1103/PhysRevD.103.106014}{\emph{Phys. Rev. D}
  {\bfseries 103} 106014 (2021)}
  [\href{https://arxiv.org/abs/2009.05016}{{\ttfamily arXiv:2009.05016}}].

\bibitem{Soltani:2021zmv}
F.~Soltani, C.~Rovelli and P.~Martin-Dussaud, \emph{{End of a black
  hole{\textquoteright}s evaporation. II.}},
  \href{https://doi.org/10.1103/PhysRevD.104.066015}{\emph{Phys. Rev. D}
  {\bfseries 104} 066015 (2021)}
  [\href{https://arxiv.org/abs/2105.06876}{{\ttfamily arXiv:2105.06876}}].

\bibitem{Christodoulou:2023psv}
M.~Christodoulou, F.~D'Ambrosio and C.~Theofilis, \emph{{Geometry transition in
  spinfoams}}, \href{https://doi.org/10.1088/1361-6382/ad6114}{\emph{Class.
  Quant. Grav.} {\bfseries 41} 195029 (2024)}
  [\href{https://arxiv.org/abs/2302.12622}{{\ttfamily arXiv:2302.12622}}].

\bibitem{Frisoni:2023agk}
P.~Frisoni, \emph{{Numerical approach to the black-to-white hole transition}},
  \href{https://doi.org/10.1103/PhysRevD.107.126012}{\emph{Phys. Rev. D}
  {\bfseries 107} 126012 (2023)}
  [\href{https://arxiv.org/abs/2304.02691}{{\ttfamily arXiv:2304.02691}}].

\bibitem{Dona:2024rdq}
P.~Dona, H.M.~Haggard, C.~Rovelli and F.~Vidotto, \emph{{Tunneling of quantum
  geometries in spinfoams}},
  \href{https://doi.org/10.1103/PhysRevD.109.106016}{\emph{Phys. Rev. D}
  {\bfseries 109} 106016 (2024)}
  [\href{https://arxiv.org/abs/2402.09038}{{\ttfamily arXiv:2402.09038}}].

\bibitem{Rovelli:2024sjl}
C.~Rovelli and F.~Vidotto, \emph{{Planck stars, White Holes, Remnants and
  Planck-mass quasi-particles. The quantum gravity phase in black holes'
  evolution and its manifestations}},
  \href{https://arxiv.org/abs/2407.09584}{{\ttfamily arXiv:2407.09584}}.

\bibitem{Han:2024rqb}
M.~Han, D.~Qu and C.~Zhang, \emph{{Spin foam amplitude of the black-to-white
  hole transition}},
  \href{https://doi.org/10.1103/PhysRevD.110.124055}{\emph{Phys. Rev. D}
  {\bfseries 110} 124055 (2024)}
  [\href{https://arxiv.org/abs/2404.02796}{{\ttfamily arXiv:2404.02796}}].

\bibitem{Dona:2025snr}
P.~Don{\`a}, H.M.~Haggard, C.~Rovelli, G.~Sreeram and J.~Taddei,
  \emph{{Spinfoam tunneling of quantum geometries in angle variables}},
  \href{https://doi.org/10.1103/68r1-pkx2}{\emph{Phys. Rev. D} {\bfseries 112}
  104004 (2025)} [\href{https://arxiv.org/abs/2507.16633}{{\ttfamily
  arXiv:2507.16633}}].

\bibitem{Perelomov:1971bd}
A.M.~Perelomov, \emph{{Coherent states for arbitrary lie groups}},
  \href{https://doi.org/10.1007/BF01645091}{\emph{Commun. Math. Phys.}
  {\bfseries 26} 222 (1972)}.

\end{thebibliography}


\end{document}